\documentclass[twocolumn]{aastex62}
\usepackage{url}
\usepackage[bookmarks=false]{hyperref}
\received{xxxx, 2019}
\revised{xxxx, 2019}
\accepted{March 29, 2019}
\submitjournal{ApJ}

\shorttitle{$z\sim 4$ proto-BCGs}
\shortauthors{Ito et al.}

\begin{document}

\title{The brightest UV-selected galaxies in protoclusters at $z\sim4$: Ancestors of Brightest Cluster Galaxies?}

\author{Kei Ito}
\email{kei.ito@nao.ac.jp}
\affiliation{Department of Astronomical Science, SOKENDAI (The Graduate University for Advanced Studies), Mitaka, Tokyo, 181-8588, Japan} 
\affiliation{Division of Optical and Infrared Astronomy, National Astronomical Observatory of Japan, Mitaka, Tokyo, 181-8588, Japan}
\author{Nobunari Kashikawa}
\affiliation{Department of Astronomy, School of Science, The University of Tokyo, 7-3-1 Hongo, Bunkyo-ku, Tokyo, 113-0033, Japan}
\affiliation{Division of Optical and Infrared Astronomy, National Astronomical Observatory of Japan, Mitaka, Tokyo, 181-8588, Japan}
\author{Jun Toshikawa}
\affiliation{Institute for Cosmic Ray Research, The University of Tokyo, 5-1-5 Kashiwa-no-Ha, Kashiwa, Chiba, 277-8582, Japan}
\author{Roderik Overzier}
\affiliation{ Observat\'{o}rio Nacional, Rua Jos\'{e} Cristino, 77. CEP 20921-400, Sa\~{o} Crist\'{o} va\~{o}, Rio de Janeiro-RJ, Brazil}
\affiliation{Institute of Astronomy, Geophysics and Atmospheric Sciences, Department of Astronomy, University of S\~{a}o Paulo, Sao Paulo, SP 05508-090, Brazil}
\author{Masayuki Tanaka}
\affiliation{Department of Astronomical Science, SOKENDAI (The Graduate University for Advanced Studies), Mitaka, Tokyo, 181-8588, Japan} 
\affiliation{Division of Optical and Infrared Astronomy, National Astronomical Observatory of Japan, Mitaka, Tokyo, 181-8588, Japan}
\author{Mariko Kubo}
\affiliation{Division of Optical and Infrared Astronomy, National Astronomical Observatory of Japan, Mitaka, Tokyo, 181-8588, Japan}
\author{Takatoshi Shibuya}
\affiliation{Kitami Institute of Technology, 165 Koen-cho, Kitami, Hokkaido 090-8507, Japan}
\author{Shogo Ishikawa}
\affiliation{Division of Theoretical Astronomy, National Astronomical Observatory of Japan, Mitaka, Tokyo, 181-8588, Japan}
\author{Masafusa Onoue}
\affiliation{Max-Planck-Institut f\"{u}r Astronomie, K\"{o}nigstuhl 17, D-69117 Heidelberg, Germany}
\author{Hisakazu Uchiyama}
\affiliation{Department of Astronomical Science, SOKENDAI (The Graduate University for Advanced Studies), Mitaka, Tokyo, 181-8588, Japan} 
\affiliation{Division of Optical and Infrared Astronomy, National Astronomical Observatory of Japan, Mitaka, Tokyo, 181-8588, Japan}
\author{Yongming Liang}
\affiliation{Department of Astronomical Science, SOKENDAI (The Graduate University for Advanced Studies), Mitaka, Tokyo, 181-8588, Japan} 
\affiliation{Division of Optical and Infrared Astronomy, National Astronomical Observatory of Japan, Mitaka, Tokyo, 181-8588, Japan}
\author{Ryo Higuchi}
\affiliation{Institute for Cosmic Ray Research, The University of Tokyo, 5-1-5 Kashiwa-no-Ha, Kashiwa, Chiba, 277-8582, Japan}
\affiliation{Department of Physics, Graduate School of Science, The University of Tokyo, 7-3-1 Hongo, Bunkyo-ku, Tokyo 113-0033, Japan}
\author{Crystal L. Martin}
\affiliation{Department of Physics, University of California, Santa Barbara, CA, 93106, USA}
\author{Chien-Hsiu Lee}
\affiliation{National Optical Astronomy Observatory, 950 North Cherry Avenue, Tucson, AZ 85719, USA}
\author{Yutaka Komiyama}
\affiliation{Division of Optical and Infrared Astronomy, National Astronomical Observatory of Japan, Mitaka, Tokyo, 181-8588, Japan}
\author{Song Huang}
\affiliation{Department of Astronomy and Astrophysics, University of California Santa Cruz, 1156 High St., Santa Cruz, CA 95064, USA}

\begin{abstract}
We present the results of a survey of the brightest UV-selected galaxies in protoclusters. These proto-brightest cluster galaxy (proto-BCG) candidates are drawn from 179 overdense regions of $g$-dropout galaxies at $z\sim4$ from the Hyper Suprime-Cam Subaru Strategic Program identified previously as good protocluster candidates. This study is the first to extend the systematic study of the progenitors of BCGs from $z\sim2$ to $z\sim4$. We carefully remove possible contaminants from foreground galaxies and, for each structure, we select the brightest galaxy that is at least 1 mag brighter than the fifth brightest galaxy. We select 63 proto-BCG candidates and compare their properties with those of galaxies in the field and those of other galaxies in overdense structures. The proto-BCG candidates and their surrounding galaxies have different rest-UV color $(i - z)$ distributions to field galaxies and other galaxies in protoclusters that do not host proto-BCGs. In addition, galaxies surrounding proto-BCGs are brighter than those in protoclusters without proto-BCGs. The image stacking analysis reveals that the average effective radius of proto-BCGs is $\sim28\%$ larger than that of field galaxies. The $i-z$ color differences suggest that proto-BCGs and their surrounding galaxies are dustier than other galaxies at $z\sim4$. These results suggest that specific environmental effects or assembly biasses have already emerged in some protoclusters as early as $z \sim 4$, and we suggest that proto-BCGs have different star formation histories than other galaxies in the same epoch.
\end{abstract}

\keywords{early Universe --- galaxies: clusters: general --- galaxies: high-redshift }
 
\section{Introduction} \label{sec:intro}
The evolution of galaxies is well known to be closely linked to their surrounding environments. An enormous number of previous studies have shown that galaxies residing in local cluster regions tend to be elliptical \citep[e.g.,][]{Dressler80}, and have higher stellar masses, lower star formation rates, and older ages \citep[e.g.,][]{Thomas05, Bamford09}. The growth of galaxies is linked to both mergers and gas accretion. Their rates are expected to be higher in the overdense regions, than in fields at high-redshifts. These overdense regions in high redshifts are called ``protoclusters''. Therefore, the overdense region at the high redshift have the possibility to represent some distinct properties than other blank field. The environmental dependence of the stellar population of galaxies appears at $z\sim2-3$ \citep[e.g.,][]{Kodama07, Kubo13}. However, we are still not sure about when the distinct characteristics of cluster galaxies emerge, and what process is responsible. \par
Brightest cluster galaxies (BCGs), which is the most massive and optically luminous galaxy in a galaxy cluster, are thought to be significantly affected by the environmental factors. The properties of local BCGs are different to those of field early-type galaxies in several aspects. For example, \citet{Bernardi07} measured the size-luminosity relationships of early-type BCGs, which were extracted from the Sloan digital sky survey (SDSS) local cluster C4 catalog by \citet{Miller:2005kc}. They found that BCGs have steeper size-luminosity gradients than early-type populations, which suggests that BCGs evolve via dry mergers with quiescent galaxies. \citet{VonderLinden:2007ev} argued that local BCGs tend to have different fundamental planes to those of elliptical galaxies. While the majority of BCGs are quiescent galaxies, just like typical massive galaxies, some local BCGs have been detected at $22\ {\rm \mu m}$, and their dust-embedded star formation rates are a few to even $\sim 100\ {\rm M_{\odot}/yr}$ \citep{Runge18}. They also host radio-loud active galactic nuclei more frequently than elliptical galaxies \citep[e.g.,][]{VonderLinden:2007ev, Best07}. \par
\citet{DeLucia07} presented a semi-analytic model of the BCG formation, which indicates that BCGs are finally formed by experiencing frequent minor mergers after $z\sim0.5$, while about $80\ (50) \%$ of the stellar mass of BCGs has already emerged from various small galaxies by $z\sim 3(5)$. \citet{Gu18} suggested a coordinated assembly of BCG components based on observations of Abell 382: their building blocks are low mass galaxies that were quenched before mergers due to the influence of dense environments. \citet{Laporte13} carried out simulations and argued that BCGs have grown in sizes by a factor of $5-10$ and in mass by a factor of $2-3$ from $z=2$ to $z=0$. To understand how BCGs form and are assembled their stellar masses, and what physical mechanisms affect their evolution, it is essential to make direct observations for progenitor BCGs (proto-BCGs) in the high-$z$ universe. 
\par So far, statistical studies of proto-BCGs have reached as far as $z\sim2$. \citet{Zhao16} selected proto-BCGs at $z\sim2$ by combining the results of observation with a semi-analytical model. They investigated the evolution of the structural parameters, stellar mass, and star formation rate and found that proto-BCGs at $z>2$ have much smaller effective radii and S\'ersic index than local BCGs do. \citet{Bonaventura17} found that the stacked far-infrared spectral energy distributions (SEDs) of their BCG sample at $0<z<1.8$ match those of a star-forming galaxy. This suggests that BCGs undergo continuous star-formation, contrary to the scenario in which BCGs passively evolve through a series of gas-poor minor mergers beyond $z\sim4$. Some researchers have recently tried to connect typical galaxies at even higher redshifts and local BCGs. \citet{Kubo16} identified one massive, quiescent and compact galaxy, as a plausible candidate of proto-BCG, in SSA22 at $z = 3.1$. They argued for a two-phase scenario of the development of BCGs, in which BCGs initially gain mass while maintaining a compact size, then increase in size through mergers. Dusty sub-millimeter galaxies and radio-loud AGNs located in overdense regions are favorable candidates for the initial states of BCGs \citep[e.g.][]{Daddi17, Miller18}. However, the nature of the primary progenitors of BCGs is not clear.\par
Although direct observations of proto-BCGs in the middle of the assemblies at high-redshift ($z\gtrsim3$) are essential for us to understand the BCG formation, these are challenging; the critical difficulty is the insufficient sample of protoclusters. Local BCGs reside in the densest environments are hosted by the most massive halos. Therefore, it is reasonable to assume that their progenitors are also likely to reside in the most overdense regions, even at high-redshifts. Due to the small number density of protoclusters, there are only confirmed $\geq10$ protoclusters at $z\sim4$ \citep{Overzier16}. Moreover, most of them have been discovered by surveying galaxies around quasars or radio galaxies \citep[e.g.,][]{Venemans07, Overzier08}, and these samples could be biased due to their unique environments, which are replete with the intense radiation. Another way to detect protoclusters is to use Ly$\alpha$ emitters or dropout galaxies to trace large-scale structures \citep[e.g.,][]{Steidel98, Lemaux14, Cucciati14, Toshikawa16, Higuchi18}. This method is a blind survey, and thus less biased than the former approach; however, it is observationally expensive to survey a wide area. In this situation, we need a large and systematic sample of protoclusters at $z>3$ so that we can determine how BCGs form, and the effects of their primordial environments.\par
To overcome this shortage of protoclusters, we have recently constructed the largest and the most systematic sample of candidate protoclusters at $z\sim4$ to date \citep{Toshikawa18}. We used the internal data release from the Hyper Suprime-Cam Subaru Strategic Program (HSC-SSP) \citep{Aihara18a}, which is the largest and the deepest optical multi-photometric survey by Subaru/HSC \citep{Miyazaki18}. Due to its large coverage field ($\sim 121\ {\rm deg^{2}}$) and depth (e.g., $\sim25.8$ mag in $i$-band at $5\sigma$), \citet{Toshikawa18} identified 179 unique protocluster candidates based on the overdensity of $g$-dropout galaxies. This large and homogeneous sample of protoclusters is ideal for systematically studying the nature of the proto-BCGs.\par
In this paper, we present the results of a survey of the brightest UV-selected cluster galaxies, which are likely proto-BCG candidates, at $z\sim 4$. The selection of UV-bright galaxies as proto-BCGs is motivated by the fact that most star-forming galaxies are found to populate the main sequence \citep{Daddi07, Song16} on the SFR-$M_{*}$, indicating that galaxies with higher star-formation rates (SFRs) are more massive. It should be noted, however, that the brightest and most massive galaxies in high-$z$ protoclusters are not always progenitors of BCGs. The majority of proto-BCGs can be submillimeter galaxies or AGNs instead of UV-bright star-forming galaxies. Furthermore, proto-BCGs are not necessarily the single brightest galaxies, and multiple high-$z$ progenitors could assemble into a single BCG at low-$z$ \citep[e.g.,][]{Ragone18}. Nevertheless, in this study, we show that such intriguing galaxy populations exist in several protoclusters at $z\sim4$, with distinct characteristics from other cluster members.\par
In Section 2, we introduce the HSC data and the protocluster catalog used in this paper. We also describe our contamination estimation method and the selection of proto-BCG candidates. In Section 3, we present a comparison between the $ i - z $ colors, which probes the UV-slope of proto-BCG candidates and field galaxies. We compare the size of the proto-BCG candidate to that of field galaxies in Section 4. In Section 5, we discuss our results in the context of the BCG evolution. Finally, we summarize the paper in Section 6. We assume that cosmological parameters are $H_{0} = 70 {\rm\  km\  s^{-1}\ Mpc^{-1}},\  \Omega_{m} = 0.3$ and $\Omega_{\Lambda} = 0.7$. We use the AB magnitude system to derive magnitudes.
\section{Data and sample selection}\label{sec:data}
\subsection{Protocluster Candidates Selection}
In this paper, we use the HSC-SSP (HSC Subaru Strategic Program) S16A internal data release \citep[][]{Aihara18b} and the catalog of protocluster candidates at $z\sim4$ constructed by  \citet{Toshikawa18}. Here, we briefly summarize the selection of galaxies and protocluster candidates. We refer the reader to \citet{Toshikawa18} for further details.  HSC-SSP is composed of three layers: Wide, Deep, and Ultra Deep. The Wide layer ($i\sim26\ {\rm mag}$ at $5\sigma$ depth) has the largest area, and it is appropriate to search protoclusters, whose number density is very low. \citet{Toshikawa18} used five separate fields (GAMA15H, HECTOMAP, VVDS, WIDE12H, XMM) from the Wide layer. HSC-SSP data is firstly analyzed on site \citep{Furusawa18}, and reduced by {\tt hscPipe} \citep{Bosch18}, which is a modified version of Large Synoptic Survey Telescope software \citep{Ivezic08, Axelrod10, Juric15}. The filter and the dewar design are described in \citep{Kawanomoto18, Komiyama18}, respectively. To identify protocluster candidates, we first select the $g$-dropout galaxies based on the color criteria defined in Equation \ref{eq:gr} - \ref{eq:ilim}. We use these equations to detect Lyman breaks in galaxies at $z\sim4$.  \citep[e.g.,][]{Ono2018, Toshikawa16,  VanderBurg2010}.
\begin{eqnarray}
1.0\ &<&\ g-r \label{eq:gr}\\
-1.0\ &< &r-i <\ 1.0\\
1.5(r-i)\ &<&\ (g-r)-0.8\\
r\ &<&\ r_{\rm lim,3\sigma}\\
i\ &<&\ i_{\rm lim,5\sigma}\label{eq:ilim}
\end{eqnarray}
$r_{\rm lim,3\sigma}\ {\rm and} \ i_{\rm lim,5\sigma}$ are the $3\sigma$ and  $5\sigma$ limiting magnitudes in the $r-$ and $i-$bands, respectively. We used the {\tt CModel} magnitude \citep{Bosch18}, which is the magnitude measured by fitting an exponential and De Vaucouleur profile to the objects. The redshift $z\sim3.8$ is the peak of the expected redshift distribution of selected $g$-dropout galaxies as shown in \citet{Ono2018}. Hereafter, we use $z\sim3.8$ as the redshift of $g$-dropout galaxies.
\par We correct and remove galactic extinctions using the extinction map published by \citet{Schlegel98}. False or wrong detections, such as cosmic rays, bad pixels, and saturated pixels are removed by using various flags \citep[see][for details]{Toshikawa18}.
\par As the density of these galaxies, we represent the overdensity significance, which is defined as,
\begin{equation}
{\rm overdensity\ significance}  = \frac{\rho - \bar{\rho}}{\sigma},
\end{equation}
where $\rho$ is the local surface number density of galaxies, and $\bar{\rho}$ and $\sigma$ are its average and standard deviation, respectively. We calculate the overdensity significance by using the fixed aperture method. We count the number of galaxies within $r <1.8$ arcmin ($\sim 0.75$ physical Mpc at $z\sim3.8$), which is the smallest size of a protocluster in this epoch \citep{Chiang:2013fs}. We distribute the apertures in a grid pattern, with intervals of one arcmin. As the depth of field varies across the survey area, we exclude the regions where the $5\sigma $ limiting magnitudes were shallower than $26.0,\ 25.5,\ {\rm and}\  25.5$ mag in the $g-,\ r-,\ {\rm and}\ i-$ bands, respectively. We also exclude apertures in which the masked region occupies $> 50\%$ of the area (e.g., around bright stars). The total sky coverage for this protocluster survey is about ${\rm 121\ deg^{2}}$. We select regions with peak significance above $ > 4 \sigma$ as candidate protoclusters. According to \citet{Toshikawa16}, 76\% of the regions that are satisfied with this definition will grow up into cluster-sized halos, where $M_{\rm halo} > 10^{14} M_{\odot}$ at $ z\sim 0$. In total, we identify 179 protocluster candidates. 
\par We have conducted the angular clustering analysis in \citet{Toshikawa18}, and we estimate that mean dark matter halo mass of selected protoclusters at $z\sim 3.8$ is about $2.3^{+0.5}_{-0.5}\times 10^{13}\ h^{-1}\ M_{\odot}$. The result of this analysis is consistent with the current $\Lambda$CDM model, and they are expected to evolve their mean halo mass into $4.1^{+0.7}_{-0.7} \times 10^{14}\ h^{-1}\ M_{\odot}$ at $z\sim0$. It should be noted that in our previous surveys about the protocluster using the same method applied to Canada-France-Hawaii Telescope Legacy Survey (CFHTLS) Deep field data \citep{Toshikawa16}, Keck/DEIMOS and Subaru/FOCAS spectroscopy revealed the success rate of this technique to be $\sim75\%$.
\subsection{Contaminant Treatment}\label{sec:cont}
Lower redshift ($0.3<z<0.6$ for $g$-dropout galaxies) galaxies and M-type dwarfs are known to contaminate the dropout galaxy samples. We also find some falsely detected objects and low-$z$ galaxies that are too large for high-$z$ galaxies in our catalog of $g$-dropout galaxies. We attempt to remove these contaminants to enhance the purity of our $g$-dropout galaxy and proto-BCG sample.
\par First, we impose additional cuts to the $r-,\ i-$bands as follows in order to exclude objects that are heavily blended. We use the {\tt blendedness\_abs\_flux}, which is the fraction of the flux of neighboring flux that affects the flux measurement of the object of interest. Note that these cuts are the same as those made by \citet{Ono2018}.
\begin{eqnarray}
{\rm \tt rblendedness\_abs\_flux}  <  0.2 \\
{\rm \tt iblendedness\_abs\_flux}  <  0.2 
\end{eqnarray}\par
Second, by applying stricter color selection criteria, we remove possible low-$z$ galaxies whose Balmer-breaks enter the $g$-band. We select objects that have the break strength greater than the lower limit and the UV-slope $\beta$ lower than the upper limit. These additional criteria are imposed because the Balmer break is generally weaker than the Lyman break and the gradient of the UV continuum is larger than that of the optical continuum \citep[c.f.][]{Bruzual03}. 
We measure the break strength and the UV slope by linearly fitting the flux density in the $i-,\ z-,\ and\ y-$bands. We use the values of the effective wavelength of each band reported by \citet{Aihara18a}. The {\tt Python} {\tt polyfit} module from the {\tt NumPy} library is used to fit the following function:
\begin{equation}
f_{\lambda,fit}(\lambda) = A\lambda^{\beta}\label{eq:fit}
\end{equation}
where A is a constant. We measure the break strength as the flux density gap between the measured flux density of the $g$-band and the value at the effective wavelength of the $g$-band, calculated using Equation \ref{eq:fit}. After that, we normalize this value by the measured $g$-band flux density. Equation \ref{eq:break} defines this. \citet{Bouwens09} estimated the UV-slope $\beta$ of dropout galaxies in the same way and found that $\sim 97\%$ of $g$-dropout galaxies have UV-slopes lower than $-0.5$. Therefore, We adopt
\begin{eqnarray}
{\rm Break} &=& \frac{f_{\lambda,fit}(\lambda_{g,cen})-f_{\lambda,g}}{f_{\lambda,g}} \label{eq:break} \\
{\rm Break} &>& 1.5\label{eq:tbreak} \\ 
\beta &<& -0.5 \label{eq:tbeta}
\end{eqnarray}
\par The thresholds are determined as follows: First, we make a high-/low-redshift reference catalog from the HSC photo-$z$ catalog called {\tt photoz\_ephor\_ab} to evaluate the contamination rate and the completeness of our sample. In the case of the high-$z$ reference sample, we select galaxies with photo-$z$ between 3.3 and 4.4, which corresponds to the redshift range of the $g$-dropout galaxies \citep{Ono2018}. At the same time, we ensure these galaxies to be satisfied with the color selection criteria and flags, which are the same criteria used to select $g$-dropout galaxies in \citet{Toshikawa18}. In the case of the low-$z$ reference sample, we select galaxies with the same selection criteria as the high-$z$ sample, but with a photo-$z$ value between 0.3 and 0.6, which corresponds to a redshift range of galaxies whose Balmer break can be misrecognized as a Lyman break. We select the high-$ z$ and low-$z$ reference samples from 36 square degrees of randomly-selected Wide-layer fields and require the magnitude of the $i$-band of the objects to be lower than $25$ mag, so that we reliably measure the break strength and UV slope $\beta$. Then, we measure $\beta$ and the break strength of each high-$z$/low-$z$ reference catalog. We calculate the contamination rate and the completeness of the sample after imposing various thresholds and identify the appropriate thresholds for minimizing the contamination rates and maximizing the completeness of the sample. The contamination rate is estimated to be $8.12\%$ (c.f. initially $21.4 \%)$ and the completeness is $83.7\%$ according to the criteria defined by Equations \ref{eq:tbreak} and \ref{eq:tbeta}. Finally, 300,692 g-dropout galaxies are obtained. Note that the density map of these selected $g$-dropout galaxies measured by the same procedure as described in \citet{Toshikawa18} is not significantly different from that of original $g$-dropout sample. \par
Next, we evaluate the contamination rates of the selected stars. We estimate the contamination rate for proto-BCGs using near-infrared (NIR) photometric data from the UKIRT Infrared Deep Sky Survey \citep[UKIDSS;][]{Lawrence07}. We match our $g$-dropout galaxies with the objects in the UKIDSS UDS DR11 based on their positions within a margin of $1\ {\rm arcsec}$ margin. We only use UKIDSS objects for which $K$-band data are available, and $g$-dropout galaxies with $z$-band magnitudes between 21.34 mag and 23.62 mag. This is the magnitude range of proto-BCGs candidates (see Section \ref{sec:BCGsel}). The $5\sigma$ limiting magnitude of the $K$-band of UKIDSS UDS DR11 is $\sim 25.3$ AB mag\footnote{https://www.nottingham.ac.uk/astronomy/UDS/data/dr11.html}, and stars have $z-K<1$; therefore we can obtain unbiased measurements of the $z-K$ color in this magnitude range. Figure  \ref{fig:gzKs} shows the colors of these objects in the $gzK$ diagram, which is a modified version of the $BzK$ diagram \citep{Daddi04, Ishikawa15}.  We assume that the $g$-dropout objects in the star sequence (below the solid line in Figure \ref{fig:gzKs}) are contamination stars. The contamination rate is evaluated to be $\sim 11 \%$. We judge the result of this estimation to be small enough to be negligible for $g$-dropout galaxies. Also, we check the location in the $gzK$ diagram of proto-BCG candidates that we selected in the following section, and none of them are located in the star sequence.
\begin{figure}[ht]
\begin{center}
\includegraphics[width=90mm,bb = 0 0 576 504]{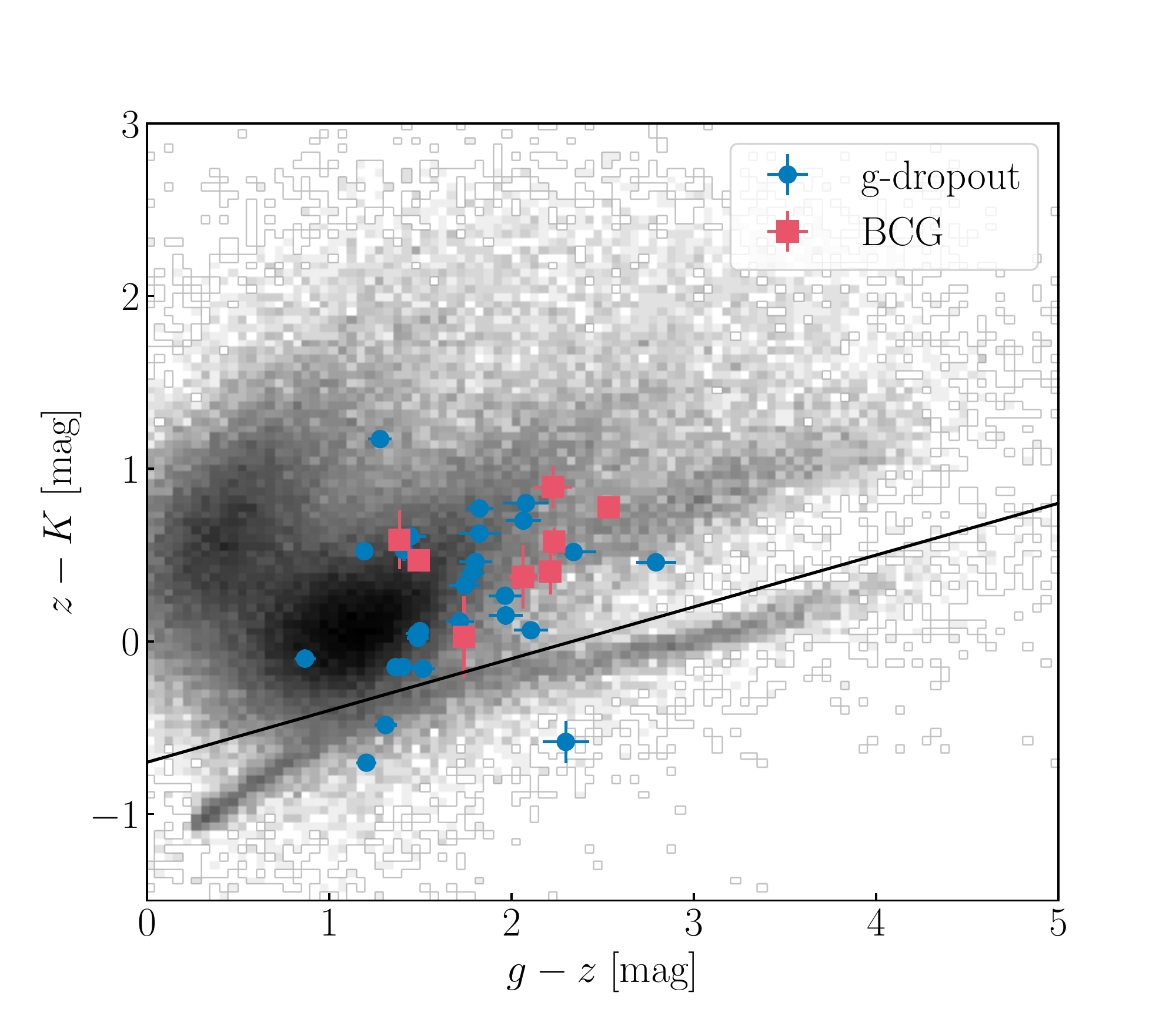} 
\caption{The $gzK$ diagram. The blue dots represent the $g$-dropout galaxies that have $K$-band photometry in UKIDSS UDS DR11 with $i$-band magnitude between 21.34 mag and 23.62 mag, which is the magnitude range of proto-BCG candidates. The gray scale shows the distribution of all sources that are detected in HSC and UKIDSS UDS. The red squares represent our proto-BCG candidates that have $K$-band photometry in UKIDSS DXS DR9 (Note that they have $K$-band photometry in DXS field, not UDS field). The star contamination rate is estimated to be $\sim 11 \%$ in $g$-dropout sample. Also, we note that none of our proto-BCGs candidates plotted in this figure are stars.}
\end{center}
\end{figure}\label{fig:gzKs}
\subsection{Proto-BCG candidates selection}\label{sec:BCGsel}
We assume that proto-BCG is the uniquely rest-UV brightest galaxy in each protocluster. To select proto-BCG candidates, we first identify protocluster members that are within $3$ arcmin of the overdensity peak of each protocluster ($1.3$ physical Mpc at $z\sim3.8$), which corresponds to the average size of the progenitor of a massive cluster at $z\sim4$ \citep{Chiang:2013fs}. It should be noted that we select members in the sky projection; hence, the contamination from fore/background field galaxies at $z\sim4$ outside the protoclusters is unavoidable. Second, we assume that the uniquely brightest galaxy compared to other galaxies in a protocluster have the more developed phase of the evolution, so in this paper we only select the brightest galaxies which are significantly brighter than other protocluster members. We measure the difference between the magnitudes of the $i$-band of the brightest member and the other galaxies in protoclusters. In this paper, we use the difference in magnitude between the brightest and the fifth brightest objects, because this difference is more significant in the case of protocluster as we show below. The distribution of the magnitude difference is plotted in Figure \ref{fig:51}. For comparison, we also measure the distribution of the magnitude difference for samples consisting of 30 randomly selected dropout galaxies from the $g$-dropout sample, which is the average number of protocluster members. The p-value of the Anderson-Darling test between the two magnitude difference distribution is $\sim 3.51\times10^{-3}$, suggesting that the null hypothesis, which is that the two distributions are the same, is rejected at a significance level of $5\%$. The protoclusters show on excess at 5th-1st magnitude $\geq1 {\rm mag}$, where uniquely bright object compared to other protocluster members can be identified. We select the brightest galaxies in protoclusters with $ \geq1{\rm\ mag}$ differences as proto-BCG candidates.
\par We identify a spectroscopically confirmed counterpart in SDSS DR12 for these candidates. It is a quasi-stellar object (QSO) at $z=4.0$. In this paper, we focus on UV-bright galaxies; therefore, we exclude this object from our analysis, and instead, we select the second brightest galaxy in the protocluster field, which is $\sim2\ {\rm mag}$ brighter than the fifth brightest galaxy. It is possible that we selected other QSOs that were not spectroscopically confirmed as proto-BCGs, but the number of such samples is expected to be low, based on the number density of QSOs. In total, we obtain 63 candidate proto-BCGs. Their brightness ranges from $21.34<m_{i}<23.68$.
\par We also note that there are nine second brightest galaxies that are 1 mag or more brighter than the fifth brightest galaxies in each protocluster. We also conduct the same procedure as we will show by using proto-BCGs sample that these second brightest galaxies are included. It suggests that even if we include these galaxies, the results do not change. Therefore, we do not include the second brightest galaxies and focus on only the brightest galaxies. 
\par We consider the possibility that these proto-BCG candidates are fore or background high-redshift galaxies. We calculate the probability that fore/background galaxies of having the same brightness as the proto-BCG candidates coincidentally. If we assume that the distribution of fore/background galaxies is random, then this probability only depends on the brightness and is equal to the fraction of galaxies with the brightness levels considered in this paper. Based on the luminosity function derived by \citet{Ono2018}, we find that the fraction of galaxies with the same brightness as the proto-BCG candidates $(21.34<m_{i}<23.68)$ out of all galaxies detected by the HSC is  $3.60\times 10^{-3}$. The average number of members of each protocluster is roughly $30$. Even if all of the objects in a protocluster are fore/background galaxies, the expected number of such galaxies is 0.11. This value is negligible; hence, we conclude that it is unlikely that proto-BCGs are fore or background galaxies.
\begin{figure}[ht]
\begin{center}
\includegraphics[width = 90 mm,bb =0 0 432 288 ]{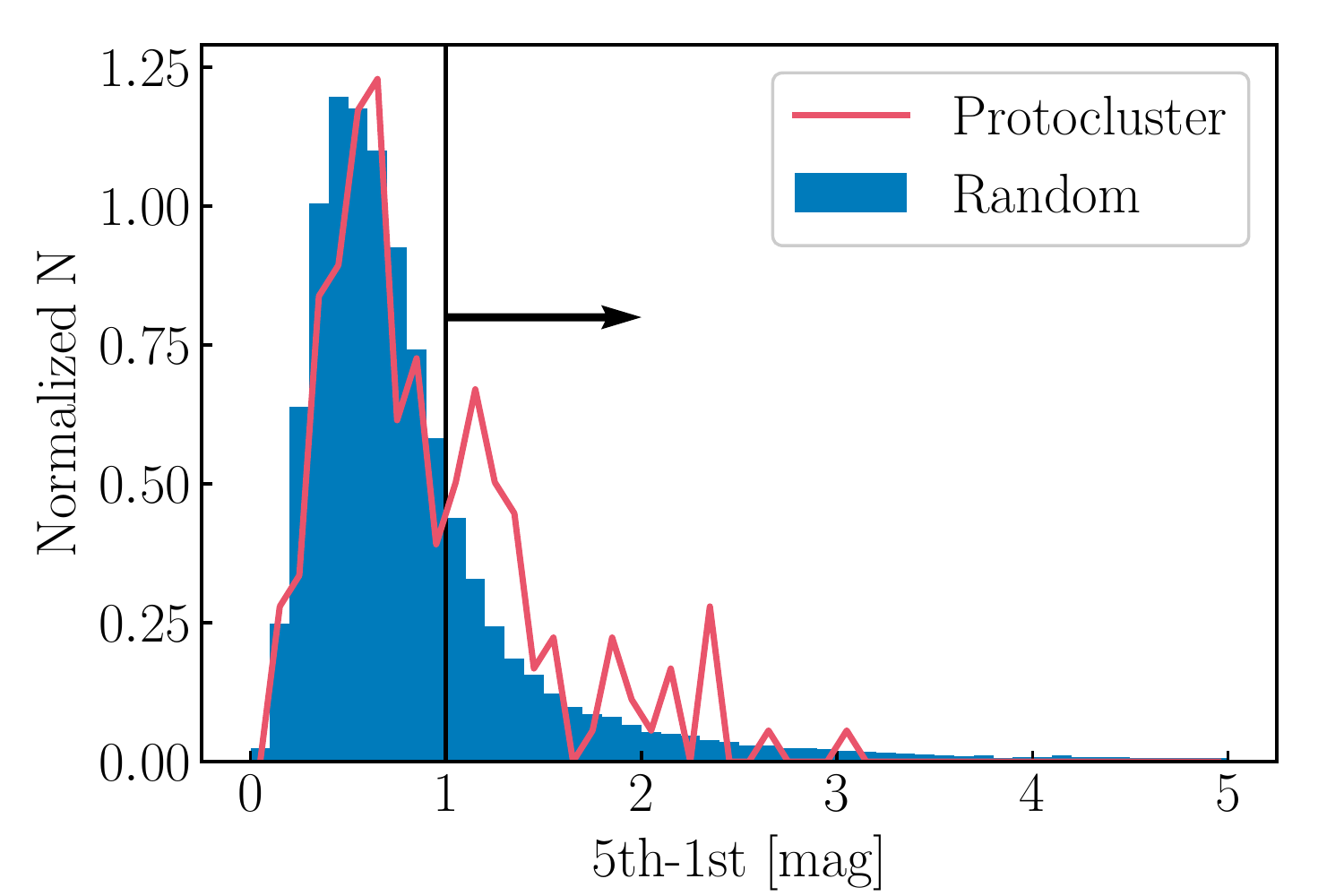}
\caption{The $i$-band magnitude difference between the fifth brightest object and the brightest object in each protocluster (red line). The blue histogram represents the randomly selected samples. The solid black line represents the threshold for selection of proto-BCGs. We select the brightest galaxy whose magnitude is one mag or much higher than the fifth brightest galaxy.}
\end{center}
\end{figure}\label{fig:51}

\section{$i-z$ color}\label{sec:color}
In this section, we compare the observed $i-z$ color, corresponding to the rest-frame UV color at $z\sim4$, which is often representative of dust attenuation \citep[e.g.,][]{Calzetti00}. We construct four subsamples from the $g$-dropout galaxies selected in Section \ref{sec:data}.\\
\begin{description}
  \item[Subsample 1] Proto-BCG candidates (63 objects)
  \item[Subsample 2] Member galaxies of protoclusters that host proto-BCG candidates (1,727 objects in 63 protocluster regions)
  \item[Subsample 3] Member galaxies of protoclusters that do not host proto-BCG candidates (3,338 objects in 116 protocluster regions)
  \item[Subsample 4] Field galaxies (295,564 objects)
\end{description}
\par Here, we define protocluster members in the same way as in Section \ref{sec:BCGsel}.  We do not include the brightest galaxies in Subsamples 2 and 3 in order to compare their characteristics with those of Subsample 1. The average numbers of galaxies in a circle with the radius of $3\ {\rm arcmin}$ are 27.4, 28.8, and 19.2 galaxies for Subsample 2, 3, and 4, respectively. We note again that Subsamples 2 and 3 are contaminated by fore- and background field galaxies because we select protocluster members based on the sky projection. Figure \ref{fig:Ncount} shows the magnitude distributions of each of the subsamples, and we can see that Subsample 2 tends to be brighter than Subsamples 3 and 4. The absolute magnitudes, $M_{\rm UV}$, are derived from the $i$-band magnitude, assuming flat rest-UV continuum and $\overline{z} = 3.8$. $i$-band corresponds to $\sim 1600$\AA\ at $z\sim3.8$. We use the $L^{*}$ value presented in \citet{VanderBurg2010}.
\par
\begin{figure}[ht]
\begin{center}
\includegraphics[width = 90 mm, bb = 0 0 576 432 ]{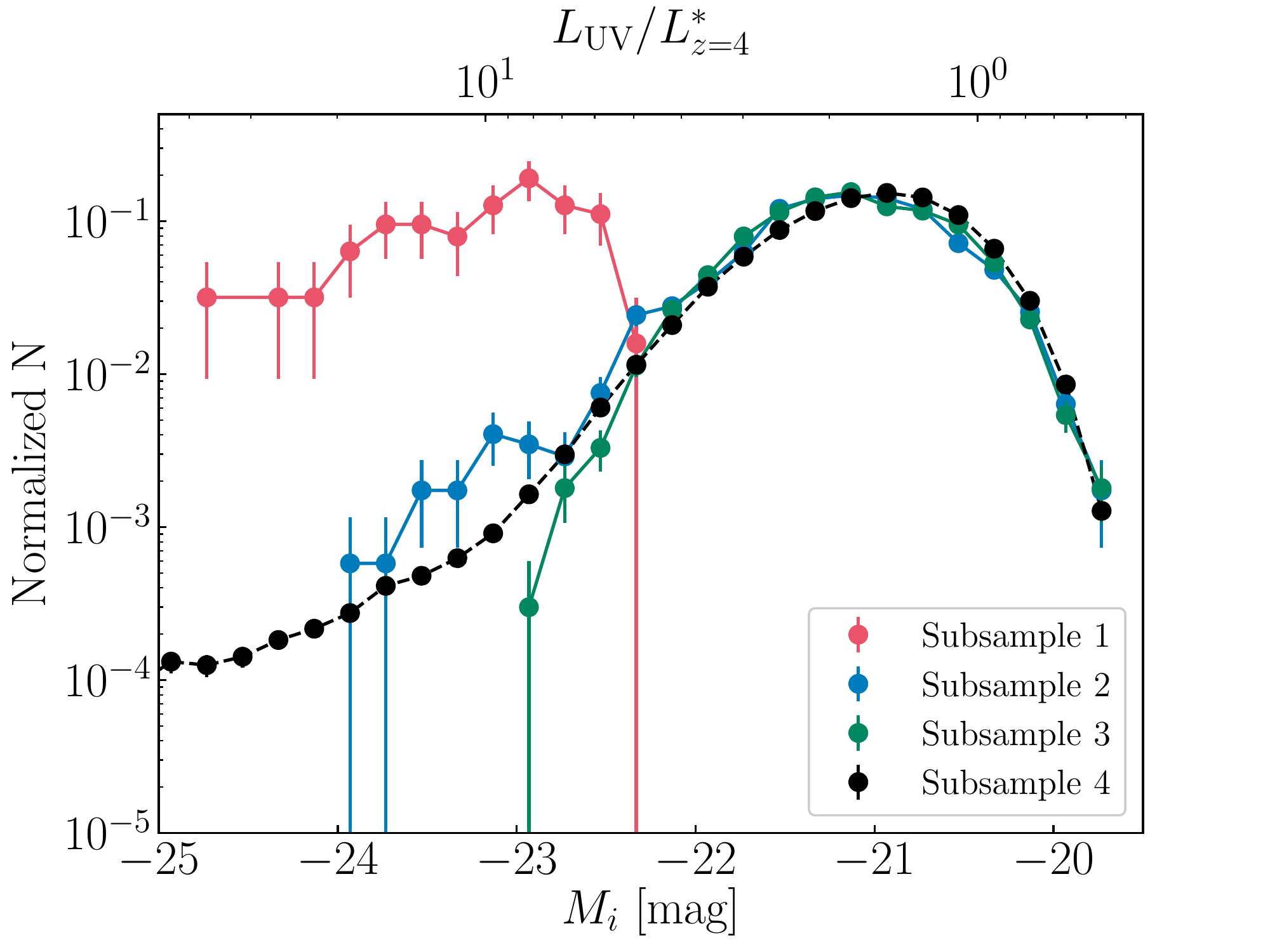}
\caption{The number count of four samples. We normalize each subsample so that the sum equals to one. The red, blue, green, and black lines show subsample 1, 2, 3, and 4, respectively. The error bar of each point shows the Poisson error.}
\end{center}
\end{figure}\label{fig:Ncount}
The UV color is known to depend on the rest-UV magnitude. \citet{Bouwens09} reported that brighter objects tend to have shallower UV-slopes (i.e., redder UV colors). This dependence is confirmed for our $g$-dropout sample, as shown in Figure \ref{fig:clmg}. Here, we only use the field galaxies brighter than $i<25.5{\rm mag}$, which is the completeness limit according to Figure \ref{fig:Ncount}. Therefore, to take into account the color-magnitude dependence, we should compare the colors of the subsamples under the same magnitude distribution. We here take the average magnitude at the middle wavelength between the $i$-band and $z$-band, as mag=$(i+z)/2$. To obtain the average color distribution of Subsample 4, we randomly select objects in each bin. We choose galaxies in Subsample 4 as many as galaxies for Subsamples whose magnitude distribution to be matched and repeated this procedure 100 times. Then, we compare the $i-z$ colors of Subsamples 1- 3 with that of the randomly selected sample  from Subsample 4. In this section, we use the {\tt Cmodel} magnitude to estimate the color and correct for the galactic extinction. 
\begin{figure}[ht]
\begin{center}
\includegraphics[width = 90 mm ,bb =  0 0 432 576 ]{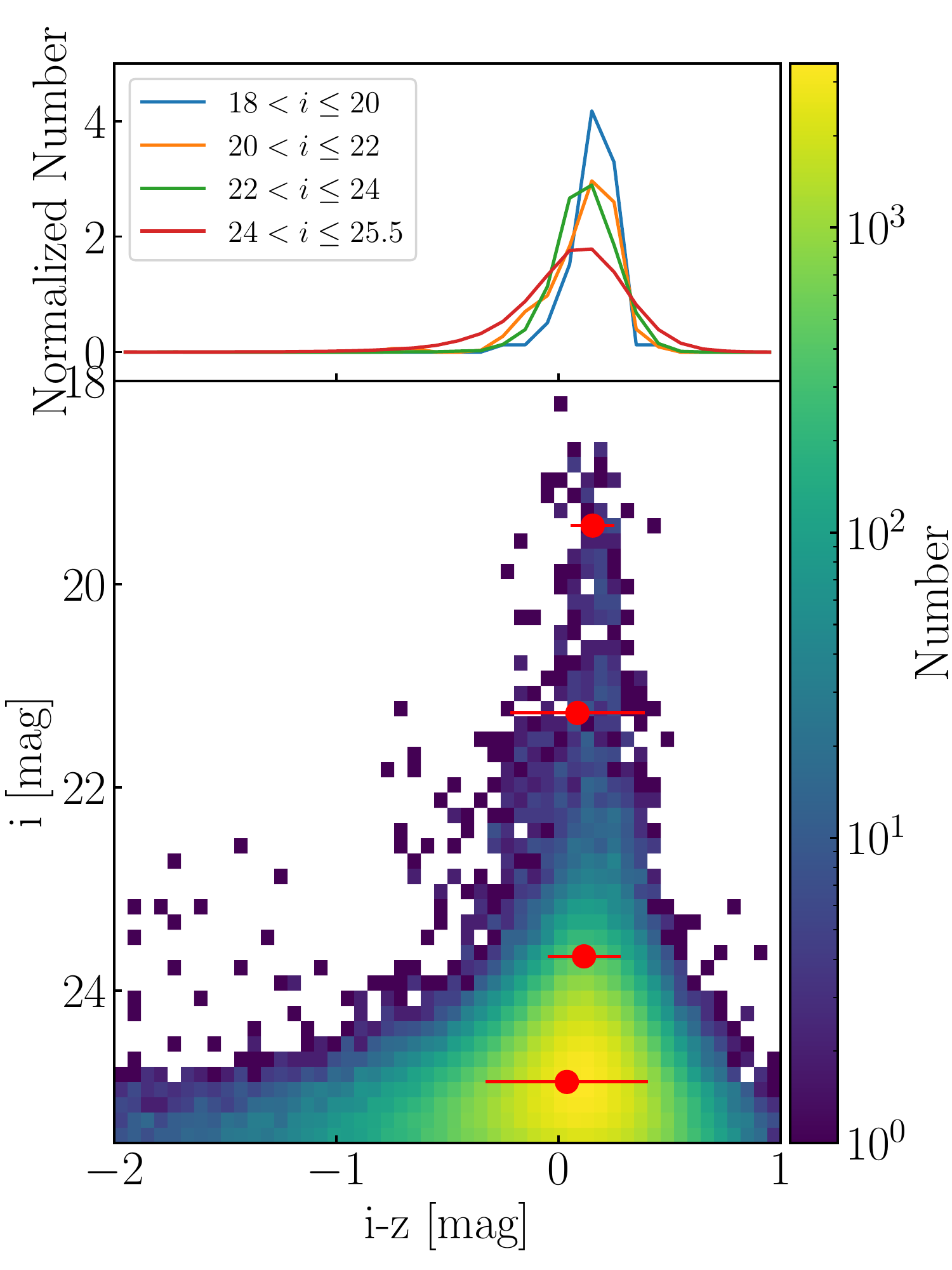}
\caption{The normalized $i-z$ color distributions of different $i$-band magnitude (upper panel) and the color-magnitude diagram (lower panel) of $g$-dropout galaxies selected in Section \ref{sec:data}. In the upper panel, we divide galaxy sample to 4 subsamples according to their $i$-band magnitude. In the lower panel, the color shows the number of galaxies. We show the mean value of each bin as red circles with error bars of $1\sigma$ standard deviation. Fainter objects tend to have more scatter of $i-z$ color, which indicates the necessity of matching the brightness.}
\end{center}
\end{figure}\label{fig:clmg}
\par First, we compare the color distribution of the proto-BCG candidates (Subsample 1) and that of field galaxies (Subsample 4). The results of this comparison are shown in the upper panel of Figure \ref{fig:color}. We match the $(i+z)/2$ magnitude distribution of the field galaxies to that of the candidate proto-BCGs. The $(i+z)/2$ magnitude range of the proto-BCGs is  $21.25\  -\  23.62 {\rm \ mag}$. The magnitude distributions of both samples are shown in the inset of Figure \ref{fig:color}. Their average $i-z$ color is $(0.1771 \pm 0.0254)\ {\rm mag}$ and $(0.1423 \pm 0.001)\ {\rm mag}$ for Subsample 1 and 4, respectively. This result shows that proto-BCG candidates are redder ($\Delta(i-z) \sim 0.03$) than field galaxies. The result of the Anderson-Darling test suggests that the p-value $p = 1.1\times10^{-2}$, so we reject the null hypothesis that these two color distributions are drawn from the same parent population at the $2\sigma$ level. Note that if we selected all brightest galaxies in each overdense regions, we do not find this statistically significant difference. However, this result alone cannot distinguish whether proto-BCGs have distinct properties, or whether all of the members of protoclusters, including proto-BCGs, tend to have different $i-z$ distributions. Therefore, we also compare member galaxies in protoclusters (Subsamples 2 and 3) to field galaxies. 
\par The middle panel of Figure \ref{fig:color} shows the results of our comparison between Subsample 2 and 4. As in the previous comparison, we match the magnitude distributions of these two subsamples. To ensure that the comparisons are fair, we only use objects in Subsample 2 that are brighter than the faintest object in Subsample 1. So we used 75 objects in Subsample 2 for this comparison. Subsample 2 is slightly redder than Subsample 4, and the Anderson-Darling test confirmed that these two samples are different ($p= 3\times 10^{-4}$) at the $2\sigma$ significance level. The average $i-z$ values of Subsample 2 and 4 are  $(0.212 \pm 0.016)\ {\rm mag}$, and $(0.154 \pm 0.0008)\ {\rm mag}$, respectively. These values also show that Subsample 2 is redder than Subsample 4, even when the $1\sigma$ error is taken into account. From these results, we conclude that members of protoclusters containing proto-BCG candidates also have redder $i-z$ colors than field galaxies.
\par Finally, we compare the $i-z$ color distribution of members of protoclusters that do not contain proto-BCG candidates (Subsample 3) with that of field galaxies (Subsample 4). We match the magnitude distribution of the field galaxies to that of Subsample 3 to eliminate any magnitude dependence of $i-z$. We only use objects brighter than the faintest object in Subsample 1. So we use 50 objects in Subsample 3 for this comparison. The bottom panel of Figure \ref{fig:color} shows the $i-z$ color distributions of these two subsamples. The Anderson-Darling test indicates that these two distributions are not different at the $2\sigma$ level ($p = 0.06$). The average $i-z$ value of two samples is $(0.1835 \pm 0.0191)\ {\rm mag}$ and $(0.1516 \pm 0.001)\ {\rm mag}$ for Subsample 3 and 4, respectively. This suggests that these subsamples have similar colors, though the average color of Subsample 3 is redder at the $1\sigma$ level. 
\par As shown in Figure \ref{fig:Ncount}, the members of Subsample 2 are brighter than those of Subsample 3. Therefore, we consider the possibility a causal relationship between the brightness and the redness of Subsample 2. We select objects in Subsample 2 with brightness levels between the brightest magnitude of Subsample 3 and the faintest magnitude of Subsample 1 and apply the same procedure as described above for comparing two samples. The average color of the objects selected from Subsample 2 is $ (0.2115 \pm 0.0193)\ {\rm mag}$ and that of Subsample 4 is $(0.1523 \pm 0.0009)\ {\rm mag}$. We also confirm that the $i-z$ color distributions of these two samples are different by carrying out an Anderson-Darling test ($p=5.5\times10^{-4}$). Therefore, we conclude that the color difference between Subsample 2 and Subsample 3 cannot be attributed to the brightness.\par
To summarize, we conclude that proto-BCGs and their surrounding galaxies are redder than field galaxies. Comparing other protocluster members without proto-BCGs to field galaxies implies that galaxies surrounding proto-BCGs are redder than other protocluster members. It is interesting to compare the $i-z$ color distributions of proto-BCG candidates and their surrounding galaxies and examine whether proto-BCG are specifically redder. However, due to the definitions of the subsamples, it is not possible to match the magnitude the distributions of these two subsamples. Also, galaxies in proto-BCG environments are brighter, as shown in Figure \ref{fig:Ncount}. Moreover, even considering their brightness, the galaxies surrounding proto-BCGs are redder than those in protoclusters that do not contain proto-BCGs.\par
\begin{figure*}[ht]
\begin{center}
\includegraphics[width = 120 mm,bb=0 0 720 1296]{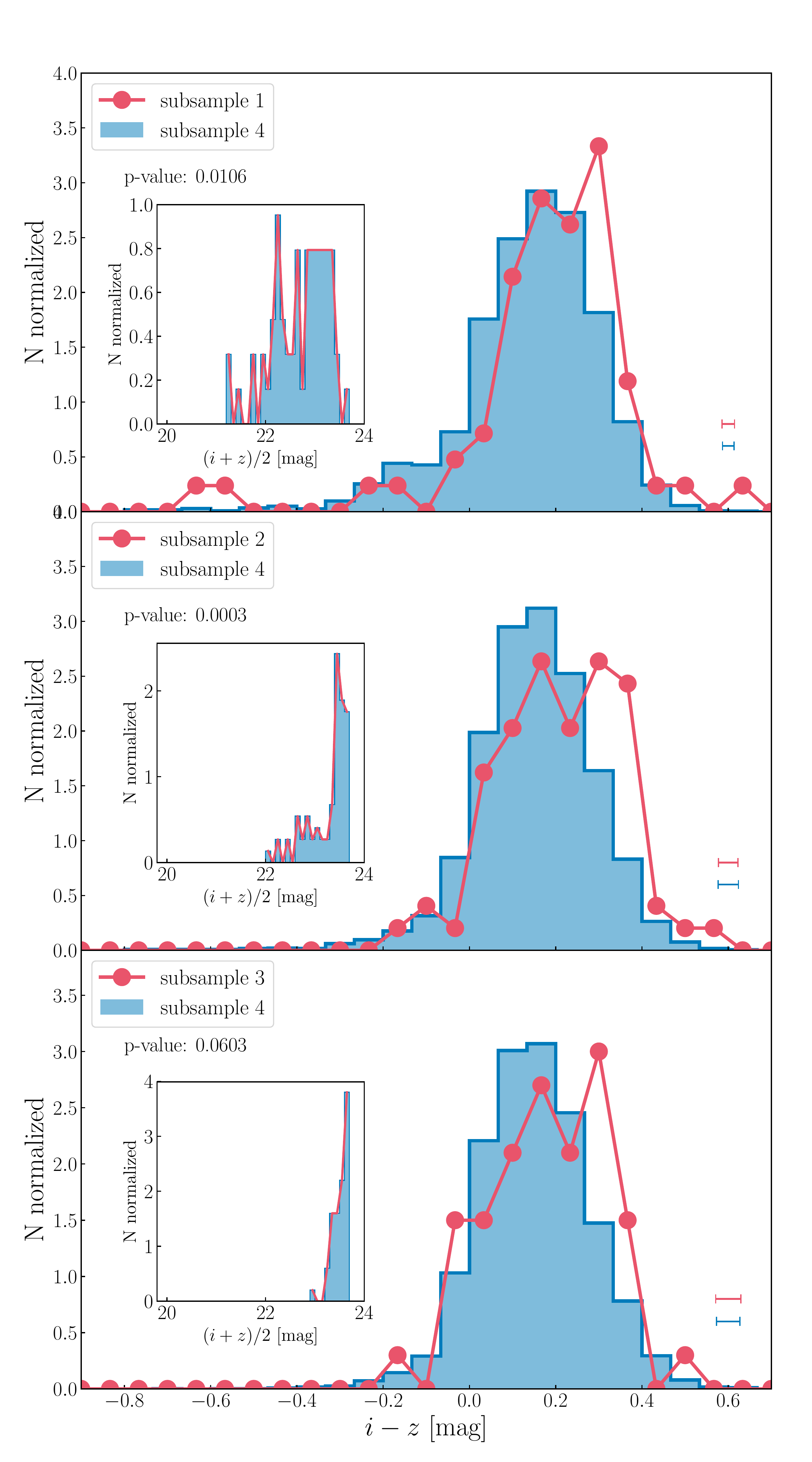}
\caption{The color distribution between magnitude-matched field galaxies and the other samples. The red line in each panel shows the color distribution of proto-BCGs (upper panel), members of protoclusters with proto-BCG candidates (middle panel), and members of protoclusters without proto-BCG candidates (bottom panel). The blue histogram in each panel represents the $i - z$ distribution of magnitude-matched field galaxies. In each inset, $(i+z)/2$ distributions of field galaxy sample (blue) and its comparison samples (red). The error bars at the lower-right corner in each panel illustrate mean uncertainties in $i-z$. }
\end{center}
\end{figure*}\label{fig:color}

\section{Size}\label{sec:mor}
The size of a galaxy is a fundamental parameter that can be used to characterize their formation history. Nearby BCGs are found to be larger than other elliptical galaxies of the same brightness \citep{Bernardi07}, which implies that many minor mergers contribute to the formation of BCGs. In this section, we compare the average sizes of the proto-BCG candidates to those of field galaxies.\par
\subsection{The stacked radial profile}
We carry out a stacking analysis to measure the average radial profile and the average size of each sample. We use the $i$-band image, which corresponds to the rest-frame $\sim 1600 {\rm \AA}$ because it provides the best images from the HSC-SSP survey strategy \citep{Aihara18a}. We select random field galaxies for the field galaxies sample, avoiding duplication, to match their $i$-band magnitude distribution with that of proto-BCG candidates in the manner described in the previous section. We repeat this process 10 times and finally, construct a field galaxy catalog of 628 galaxies. We use the same stacking method as reported by \citet{Momose14}. In brief, the procedures are:\\
 (1) Image cutouts\\
We generate postage stamps in the $i$-band with a size of $8 {\rm \  arcsec}\times 8 {\rm \  arcsec}$, which corresponds to $3.35\times10^{3}{\rm \ physical\ pc}^{2}$ at $z \sim 3.8$.\\
 (2) PSF matching\\
 We obtain point spread function (PSF) images, then measure the full width at half maximum (FWHM) of each PSF while approximating each PSF as a Gaussian. We smooth all of the images to $0.806$ arcsec, which is the lowest resolution obtained.\\
 (3) Normalization\\
  To avoid weighting brighter objects, we normalize each image to the peak count of object.\\
 (4) Stacking\\
 We stack the images using the {\tt Imcombine} task from the {\tt IRAF} package and apply the average stacking. Following \citet{Momose14}, we applied $3\sigma$ clipping to remove unusually bright pixels. The central position of each object was based on the HSC catalog.\par
The radial profiles of the stacked images are shown in Figure \ref{fig:radpl}. We bin the points in 0.2 arcsec bins. The measurement of the radial profile error is summarized in the Appendix. To make a fair comparison, we normalize the radial profile at the center of each image. Figure \ref{fig:radpl} also shows the ratio of the normalized fluxes of these two samples in the bottom panel. From the center of the object to 1.6 arcsec, we can see a moderate enhancement of proto-BCGs over $1\sigma$, suggesting that proto-BCGs have slightly more extended radial profiles than field galaxies. Note that, though the flux ratio is lower than 1, which means that proto-BCG candidates have smaller counts than field galaxies at $r >2.2\ {\rm arcsec}$, it is still within the  $1\sigma$ error. We investigate the effects of imperfections in PSF matching by smoothing the PSF of each object image and stacking all the PSFs in each subsample in the way described above. The radial profiles of the stacked PSFs of proto-BCGs and field galaxies are plotted with red and blue dashed lines in Figure \ref{fig:radpl}, respectively. There is no significant difference between the stacked PSFs. Therefore, the differences between the radial profiles of proto-BCGs and field galaxies are not mainly due to imperfections in PSF matching. Also note that if we select all brightest galaxies in each overdense regions instead of imposing the criteria ${\rm 5th-1st >1}$ mag in selecting proto-BCGs, the difference of the radial profile gets smaller. \par
\subsection{The size measurement}
We measure the effective radii of the stacked images so that we could compare our results with those of a previous study. We use {\tt GALFIT} \citep{Peng02, Peng10} to fit the two-dimensional surface brightness profile. We set the fitting model to be the same as \citet{Shibuya:2015vh} (hereafter S15). S15 measured the size distribution of the dropout galaxies at $z\sim4$ by fitting the S\'ersic profile \citep{Sersic63} to HST images and argued that star-forming galaxies have a mean S\'ersic index of 1.5 and that their effective radii are mostly unaffected by varying the S\'ersic index. We thus set n=1.5. For the test of this analysis, we set $n = 1,\ 1.5,\ 2,\ 3,\ 4, {\rm and}\  5$ and derive effective radii. The standard deviation of each effective radius is $\delta r_{e,{\rm BCG}} \sim 0.05\ {\rm kpc},\ \delta  r_{e,{\rm Field gal.}} \sim 0.09\ {\rm kpc}$, respectively. We convert angular distances to physical scales by assuming that the redshift of the objects is $z\sim3.8$. Even we consider these difference, our result does not significantly change. Therefore, we use $n=1.5$ so that we could compare our results to those reported in S15.
\par The fitting result for the stacked proto-BCG image is plotted in Figure \ref{fig:fit}. There are an $\sim 3.6 \%$ oversubtraction at the center of the image, which is seen in the right panel of the Figure \ref{fig:fit}. We calculate the effective radius $r_{e}$, by converting the effective radius along the semi-major axis $r_{e,{\rm major}}$ through $r_{e} \equiv  r_{e,{\rm major}} \sqrt{q}$, where $q$ is the axis ratio of the object. We estimate the errors in the effective radii of these stacked images using the following procedure. First, we make an image of Gaussian random noise equivalent to a $1\sigma$ error in the radial profile, and then repeat this procedure 1,000 times. Second, we apply {\tt GALFIT} to each image and obtain the effective radius distribution. Finally, we use that the average value of this distribution as the typical value of the effective radius and its ${\rm 16^{th}/84^{th}}$ percentile as the error of the effective radius due to the uncertainty of the stacked image. We obtain an effective proto-BCG radius of $r_{e,\ {\rm BCG}} =  2.042^{+0.012}_{-0.013}{\rm \ kpc}$ and that of field galaxies of $r_{e,\ {\rm Field}} = 1.597^{+0.003} _{-0.003}{\rm \ kpc}$. We find that the effective radius of the proto-BCG candidates is slightly larger than that of the field galaxies. 
\par We estimate the uncertainty due to the resolution limit of the HSC images as follows. First, we make a mock image whose surface brightness follows S\'ersic profile with the Poisson noise. We set the spatial resolution to be the same as that of the HSC image, and the brightness to be the same as that of the stacked proto-BCG image. We set the effective radius to be the same as the size of stacked proto-BCGs, and S\'ersic index $n=1.5$. The position angle is equal to zero. Second, we smooth the profile to the worst PSF that we match in the stacking analysis. We use observational PSF image. Then, we add a sky noise to the PSF-convolved image. Finally, we apply {\tt GALFIT} to this mock image with a fixed S\'ersic index. We estimate the effective radius of this profile as $r_{e}=  1.58\pm0.16\ {\rm kpc}$. The underestimate of the effective radius does not change, even if we set the model's effective radius of the model to be the same as that of the field galaxies. This means that {\tt GALFIT} can lead to underestimates ($\sim 22\pm8\%$) of the effective radius for the HSC images. Therefore, our result of the size is the lower limit.
\par We compare these effective radii to the rest-UV size-luminosity relationships of the dropout galaxies at $z\sim4$ from S15. Figure \ref{fig:s15} shows a comparison of our results to those reported in S15. The effective radius of our stacked $g$-dropout galaxy sample is consistent with the size-luminosity relation of S15. Hence, our measurement of the effective radius of the field galaxies is consistent with S15, and the size of the proto-BCG candidates are slightly larger ($\sim28\%$).
\begin{figure}[ht]
\begin{center}
\includegraphics[width = 90 mm, bb =  0 0 576 720 ]{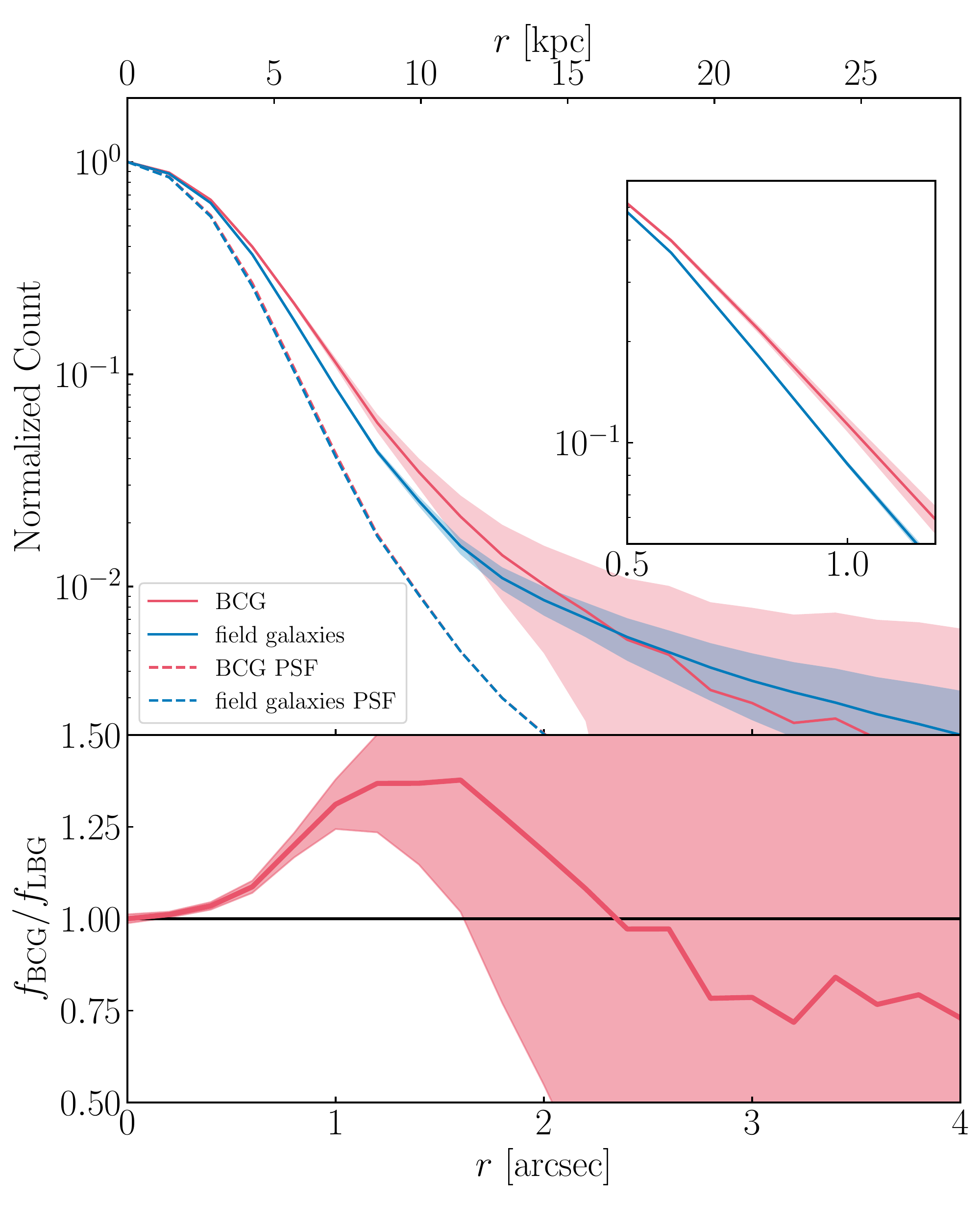}
\caption{(Upper panel) The radial profile of proto-BCG candidates (red line) and field galaxies (blue line). Shed area represents each 1$\sigma$ error. we calculate the error from sigma image of stacking produced by {\tt Imcombine}. The inset is a close-up of the small-scale range at $0.5 < r\ {\rm [arcsec]}<1.2$. We can see the difference of the radial profile more clearly. (Lower panel) The radial profile of normalized flux ratio of proto-BCG candidates and field galaxies. The red and blue dashed lines are the stacked PSF radial profile of proto-BCG candidates and field galaxies, respectively.}
\end{center}
\end{figure}\label{fig:radpl}
\begin{figure}[ht]
\begin{center}
\includegraphics[width= 90mm, bb = 0 0 864 288]{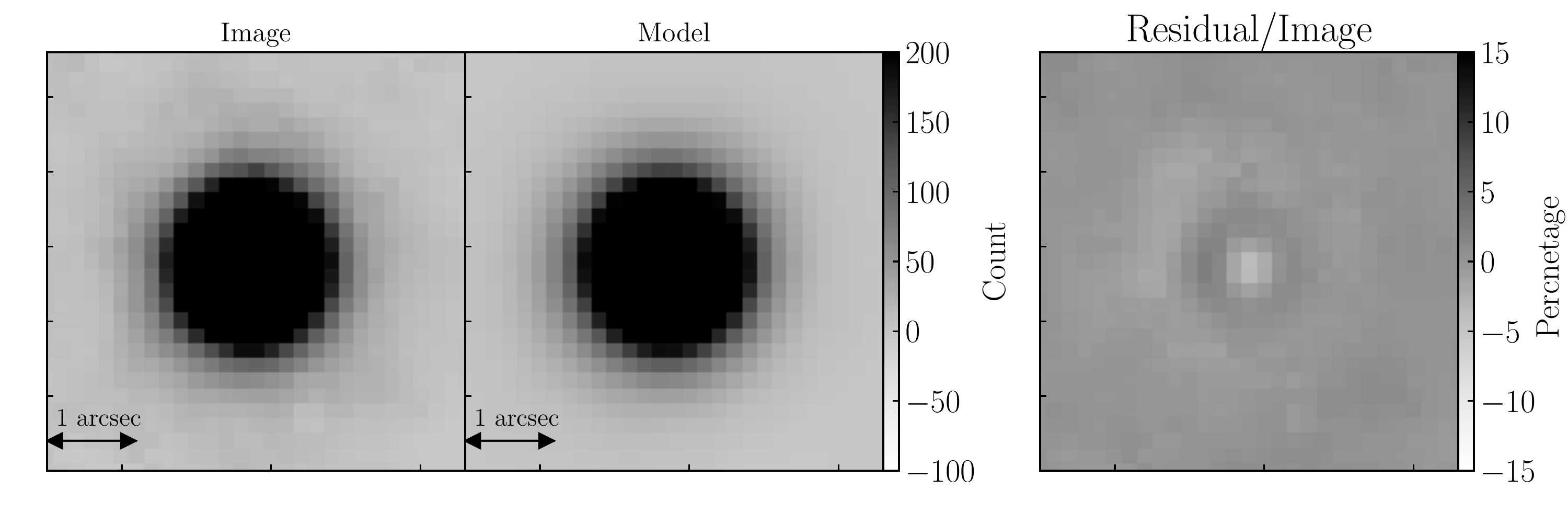}
\caption{(Left Panel) 63 proto-BCG candidates stacked image in $i$-band. (Middle Panel) The model image of the stacked image is generated by {\tt GALFIT}. (Right Panel) Residual image from the model. The contour of the right panel shows the percentage of the residual to the maximum value of the image count, while the contour of the left and middle panel shows the counts.}
\end{center}
\end{figure}\label{fig:fit}
\begin{figure}
\begin{center}
\includegraphics[width = 90 mm, bb = 0 0 576 432]{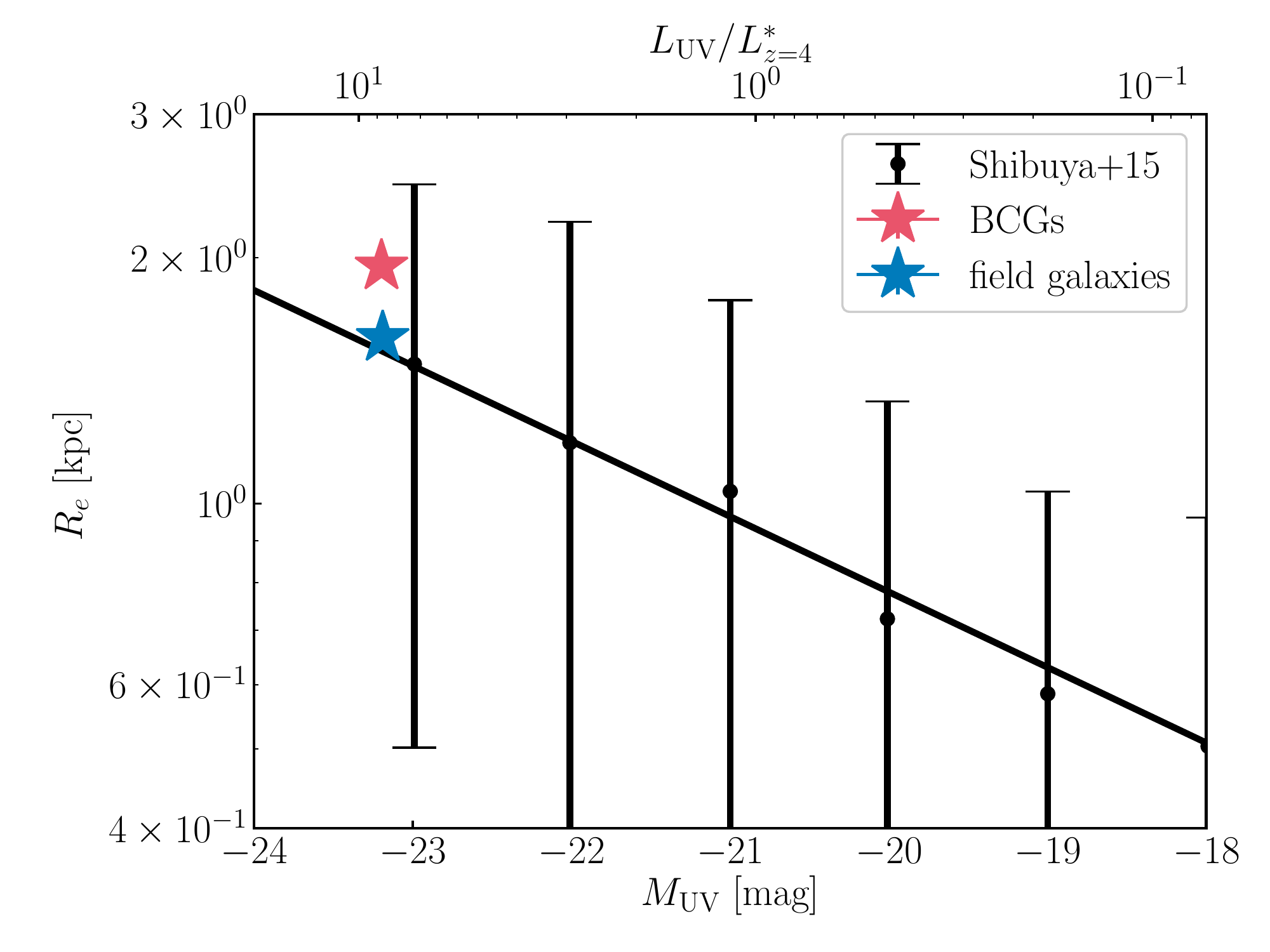}
\caption{The comparison of effective radii of the stacked image of proto-BCG candidates and field galaxies to \citet{Shibuya:2015vh}. The red star represents the effective radius of proto-BCG candidates, while the blue star represents that of field galaxies sample. The solid black line and the points with error bars is the size-luminosity relation of the distribution of dropout galaxies at $z\sim4$ from \citet{Shibuya:2015vh}. The error bars is the $16^{th}$ and $84^{th}$ percentiles of the effective radius of galaxies.}
\end{center}
\end{figure}\label{fig:s15}

\section{Discussion}\label{sec:dis}
\subsection{The difference between UV-color, and its implication}
According to the results reported in Section \ref{sec:color}, our proto-BCG candidates and their surrounding galaxies tend to be redder than field galaxies in the rest-UV. More interestingly, the members of the protoclusters that do not contain proto-BCGs are not significantly redder color in $(i-z)$ than field galaxies. 
\par The redder rest-UV color can be occurred by the dust enrichment, older age, or the enhancement of the metallicity. \citet{Bouwens09} use the SED model of $U$-dropout galaxies at $z\sim2.5$ assuming Saltpeter IMF, and they investigate the effect of several properties of galaxies to the value of its UV-slope $\beta$ (see their Figure 7). They argue that the amount of dust is the most effective to the change of $\beta$. Our proto-BCGs candidates are $\sim0.03{\rm\ mag}$ redder than other field galaxies, and it corresponds to $\Delta\beta\sim 0.3$ according to the conversion equation between $i-z$ and $\beta$ in \citet{Overzier08}. Assuming the relationship between the change of $\beta$ and that of other properties is the same for $z\sim3.8$ $g$-dropout galaxies, proto-BCGs have to be $\sim0.9{\rm \ dex}$ older than field galaxies on average if the age difference is the only cause for the color difference. Our proto-BCGs and field galaxies are Lyman break galaxies, which are generally young galaxies, so it is unlikely that proto-BCGs have such older age in general. Also, the difference of the metallicity needs to be greater than that of the case for the age in order to explain such UV color difference. Therefore, We can interpret the dust enrichment causes the redder color in the rest-UV frame of proto-BCGs.\par
Supernovae are the predominant cause of the enhancement of the dust \citep{Indebetouw13}, especially type-II supernovae \citep{Todini01}, which are caused by massive stars. This dust enrichment implies that there are more massive stars in galaxies in and around proto-BCGs than in other galaxies. The excess of UV-bright galaxies in protoclusters hosting proto-BCGs is also shown in Figure \ref{fig:Ncount}, and this suggests that the star formation activity is underway in the regions around the proto-BCG candidates. Interestingly, the star formation rate ($\sim4800 \ M_{*}/{\rm yr^{-1}}$) estimated from the formula in \citet{Kennicutt98} and the UV luminosity of our proto-BCG sample lies on the extension of the SFR evolution of low-$z$ BCGs at $0<z<1.8$ \citep{Webb15}. This suggests that a rapid increase in star formation in the central galaxies of these clusters continues until at least $z\sim4$.
 \par We suggest two scenarios for explaining the enhanced dust extinction. One is that the star formation have continued since the earlier period. In this case, the redder color of proto-BCG candidates could partially be due to the older age of the galaxies. The other is that a starburst phase occurred during the star-formation period. The starburst activity in proto-BCGs can produce more massive stars, which increase the amount of dust in proto-BCGs. These periods of starburst activity can be caused by mergers, which can occur more frequently in large overdense regions. \citet{Hine16} found that the merger fraction in the SSA22 field, which is one of the most overdense regions at $z=3.1$, is higher than that of field galaxies at the same redshift. If this tendency is ubiquitous in other overdense regions around at $z\sim4$, then the enhanced star formation may be expected to be caused by the gas supplied by merger in protocluster regions.
 \par We cannot yet reject either of these two scenarios so far. However, in any case, the results of this study suggest that proto-BCGs and their surrounding protocluster members are located in unique regions and have star formation histories distinct from those of other star-forming galaxies at $z\sim4$. Interestingly, the average overdensity significance peak of protoclusters that contain proto-BCG candidates is $ (5.068 \pm 0.149) \sigma$, whereas that of protoclusters that do not contain proto-BCG candidates is $(4.767 \pm 0.069)\sigma$. This slight difference may suggest that proto-BCG candidates are likely to be located at slightly more massive halo with more mature structure formation.
 \subsection{The effect of the dust on the size}\label{sec:mordust}
The results reported in Section \ref{sec:mor} indicate that our proto-BCG candidates are larger than other galaxies of the same brightness. These results can be closely correlated with the different UV colors. Size-luminosity relations at various redshifts and wavelengths (rest-UV, optical) \citep[e.g.,][]{Barden05, Shibuya:2015vh} indicated that brighter objects have larger sizes. In Section \ref{sec:mor}, we stacked two samples after matching the magnitude distributions. However, if the redder color is due to the dust, then the intrinsic luminosities of the proto-BCG candidates may be higher than observed due to the relatively high dust extinction. This would explain the larger sizes of candidate proto-BCGs. The dust gradients of the galaxies may also explain the difference in their size, but due to the resolution limit, we did not consider this effect in this study. \par
We investigate the effect of dust on the size of two samples by matching their dust corrected magnitude distributions. First, we construct a field galaxies sample whose dust corrected magnitude distributions matched those of the candidate proto-BCG sample. We then derive the dust obscuration $A(0.16)$ from the measured UV slope $\beta$ by using the equations proposed by \citet{Calzetti00}:
\begin{equation}
A(0.16) = 2.31(\beta-\beta_{0}),
\end{equation}
where $\beta_{0}$ is the intrinsic UV spectral slope and $\beta_{0} = -2.1$ when $\beta > -1.4$ or $\beta_{0} = -2.35$ when $\beta < -1.4$. As described in Section \ref{sec:cont}, we derive the UV slope $\beta$ of our sample based on the photometric data of the HSC $i-,\ z-,\ y$-bands. The median dust extinction of the proto-BCG candidates is $A(0.16)=2.29 {\rm\ mag}$. This value is consistent with those reported in previous studies of Lyman break galaxies at $z\sim 3 $ \citep[e.g.,][]{Meurer99}. After correcting for the dust extinction, we match the intrinsic brightness of both samples, as done in Section \ref{sec:color}. Finally, we stack the images of the two samples using the method described in Section \ref{sec:mor}. 
\par Figure \ref{fig:radpldust} shows the radial profile of these two stacked images. The difference between the radial profile of proto-BCG candidates and field galaxies remains. This result suggests that the difference of the rest-UV size between the proto-BCG candidates and field galaxies is due not only to the difference of the dust but also to other physics. There are two possibilities for explaining the difference instead of the intrinsic brightness difference. One is the concentration of the dust in the center of the galaxies. This makes the profile flatter, leading to the larger size of its profile. The other is the enhancement of the hidden satellite galaxies around proto-BCGs. From Figure \ref{fig:radpl}, the difference of the radial profiles of proto-BCG candidates and field galaxies appears at the maximum at $r\sim1\ {\rm arcsec}\ (r\sim 7\ {\rm kpc})$, which is larger than the effective radius. Therefore, more satellite galaxies around proto-BCGs make the proto-BCG's radial profile larger. 
\par \citet{Bowler17} obtained a size-luminosity relation of Lyman-break galaxies at $z\sim7$ and found there is a curvature at the brightest end ($M_{UV}\sim-23.0$) in their size-luminosity relation. They argue that it implies that the galaxy merger makes such bright galaxies. Our proto-BCGs do not have different size-luminosity relation apart from $66\%$ range of a size-luminosity relation from \citet{Shibuya:2015vh}; therefore the larger size of proto-BCGs is more preferable to be explained by causes that we suggest above. However, note that these are the just one possibility and we cannot prove this scenario yet.  
\begin{figure}[h]
\begin{center}
\includegraphics[width = 90 mm, bb =  0 0 576 720 ]{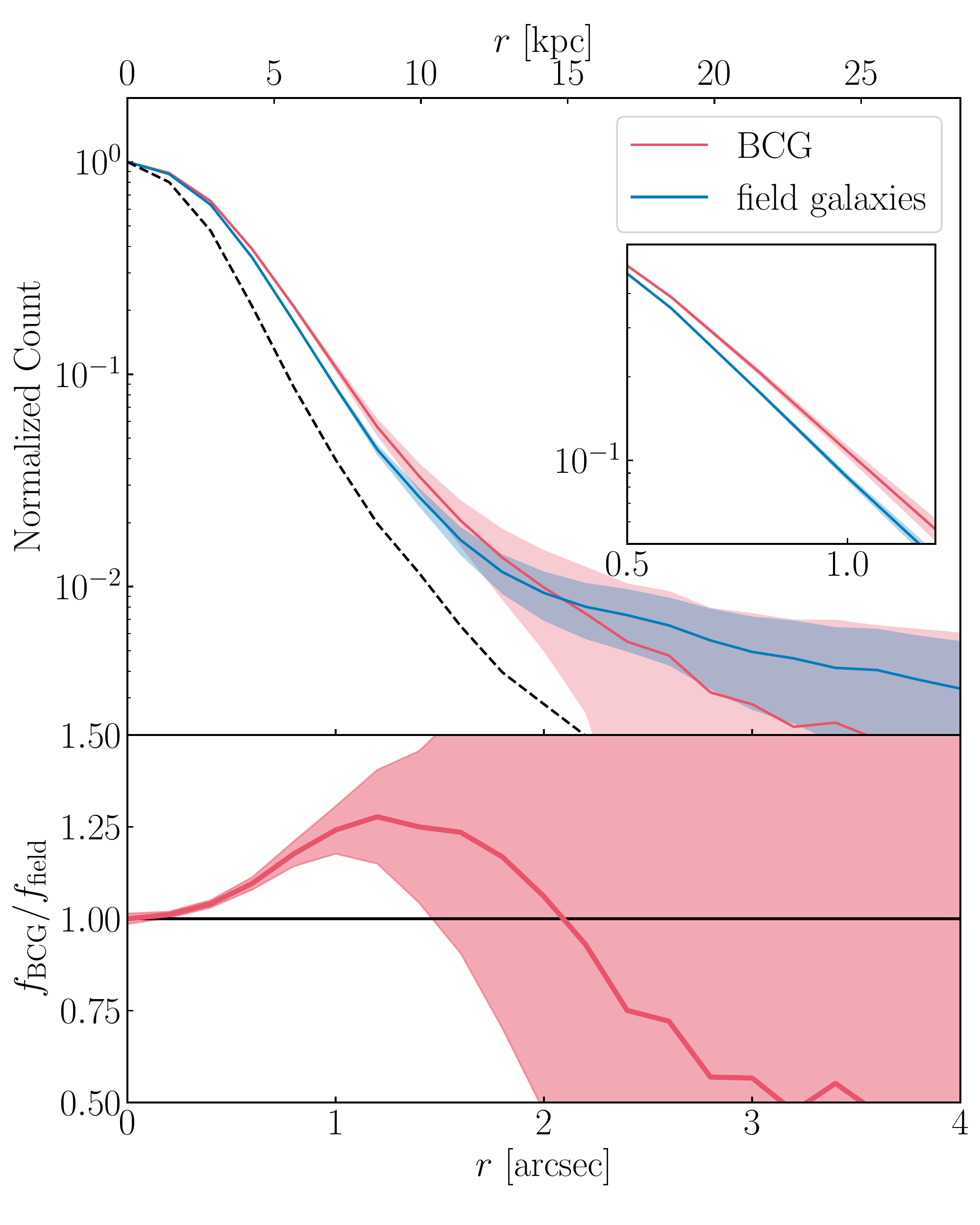}
\caption{The same as Figure \ref{fig:radpl}, but two samples for stacking were brightness-matched after correcting dust obscuration. (Upper panel) The radial profile of proto-BCG candidates (red line) and field galaxies (blue line). The black dashed line is the radial profile of PSF for the stacked images. Shed area represents each 1$\sigma$ error.  (Lower panel) The radial profile of normalized flux ratio of proto-BCG candidates and field galaxies.}
\end{center}
\end{figure}\label{fig:radpldust}

\subsection{The evolution of (proto-)BCGs}
Here, we compare these results described so far to those of other proto-BCG at $z\sim4$. \citet{Overzier08} found a protocluster at $z\sim4$ around a radio galaxy called TN J1338-1942 and suggested that this radio galaxy is likely to be a proto-BCG. Its rest-UV absolute magnitude is $\sim -23.0$ mag, and the $i-z$ color is $0.1$ mag. These values are in good agreement with our proto-BCG sample ($\langle M_{\rm UV} \rangle \sim -23.2\ {\rm  mag}$, $\langle i-z\rangle = 0.17\ {\rm  mag}$). They estimated the effective radius from the $z$-band as $R_{e}\sim4.3\ {\rm kpc}$. Our proto-BCGs have $R_{e}\sim2.04\ {\rm kpc}$, so TN J1338 is larger than our proto-BCG, possibly due to the radio jet. Even though these two have different radii, they are both larger than typical field galaxies. \par
Next, we compared the sizes of these proto-BCGs to those of BCGs at different redshifts. Size measurements of local BCGs are often based on rest-optical band images so that stellar emissions can be traced, but we measured the size in the rest-UV frame, where the flux is dominated by young stars. One may object that the results of our size measurements cannot be compared directly to those of previous studies. However, \citet{Papovich05} suggested that morphologies of galaxies at high-redshifts do not depend on their wavelengths. \citet{Shibuya:2015vh} compared the UV and optical sizes of star-forming galaxies at $1\leq z\leq3$ and found their median sizes to be comparable. Assuming that this trend holds at higher redshifts, which means at $z>3$, we compare our results to the sizes of BCGs at lower redshifts. \par
We also derive the stellar masses based on the average UV luminosity of proto-BCGs. \citet{Song16} derived a $M_{*}$-$M_{UV}$ relation in the magnitude range of $-23<M_{UV}<-16$. The average absolute magnitude of our proto-BCGs is $M_{UV} = -23.20$. Extrapolating this relation to the brighter magnitudes based on the assumption that this relationship does not flatten at the bright end, we obtain $\log{M_{*}/M_{\odot}}=10.87$. We estimate the stellar mass using $M_{*}$-$M_{UV}$ (SFR) relations from other papers \citep[][]{Gonzalez14, Speagle14} and find that masses to be consistent within $\Delta \log{M_{*}/M_{\odot}}\sim 0.2$. This value is almost the same as TN J1338 ($\sim 10^{11} M_{\odot}$) \citep{Overzier08}. \par
Figure \ref{fig:sizeevo} shows the size evolution of BCGs at the stellar masses of $10^{11}<M_{*}/M_{\odot}<10^{11.5}$, which is consistent with the same stellar mass range of our proto-BCGs. \citet{Zhao15} used the BCG catalog published by \citet{VonderLinden:2007ev} and derived a size-stellar mass relation at $0.02<z<0.1$. \citet{Furnell18} used BCGs at $0.05<z< 0.3$ in X-ray detected clusters from the spectroscopic identification of eROSITA sources (SPIDERS) survey and measured their size based on $g$-band images obtained from the SDSS. \citet{Zhao16} selected progenitors of BCGs at $z\sim2$ based on a semi-analytical model and measured their sizes in HST F160W. All of these size measurements were made in the rest-frame $\sim 5,000{\rm \AA}$ and conducted by {\tt GALFIT}. We can see that the size increases monotonically at lower redshifts. We also compared our results with those on the size evolution of massive quiescent galaxies at the same stellar mass reported by \citet{Kubo18}; they derived the size evolutional track by fitting of the sizes of the massive quiescent galaxies at $z\sim4$ and those calculated in previous studies \citep{Kubo16, Straatman15, vanderWel14}. Although there are discrepancies between \citet{Zhao15} and \citet{Furnell18}, the size evolution tracks of BCGs are above that of quiescent massive galaxies. The shape of this evolution tracks is consistent with that of general massive quiescent galaxies. On the other hand, the size evolution of BCGs differs from that of star-forming galaxies \citep{vanderWel14}. These results also imply that progenitors of BCGs have experienced different star-formation histories to star-forming galaxies and that they may have undergone earlier star-formation than massive galaxies. Some recent studies show that star-forming galaxies once experienced starburst and increase the stellar density at the center, leading to get small size \citep[e.g.,][]{Barro17, Toft14}. We compare (proto-)BCGs at various redshifts, regardless of whether star formation was active. The scenario described above indicates that (proto-)BCGs can have smaller radii than suggested by the trend shown in Figure \ref{fig:sizeevo}.\par
Figure \ref{fig:sizemass} shows a comparison between the size-stellar mass growth of BCGs and massive galaxies reported in previous papers. Compared to the evolution tracks of massive quiescent galaxies \citep{Kubo18}, BCGs are shifted towards larger sizes. Also, according to a simple toy model, the mass and size growth by the major merger is followed by $r_{e} \propto  M_{*}$, while the growth by the minor merger is followed by $r_{e} \propto M_{*}^{2}$ \citep{Bezanson09, Naab09}. We move both models track in order to overlap our data and find the fit line of the minor-merger schema to be in good agreement with the results reported by \citet{Zhao15, Furnell18}. If our proto-BCGs evolve exclusively by minor mergers with local BCGs, we expect that our proto-BCG will evolve into local BCGs with $\sim 3-4\times10^{11}\ M_{\odot}$ in \citet{Zhao15}.\par
 It should be noted, however, that each study compared are based on different BCG selection criteria. In particular, \citet{Zhao16} selected proto-BCGs based on their stellar mass and environmental density, which is different to our selection criteria. They assumed that the most massive galaxy in every overdense region is a proto-BCG, while our study, we only identify proto-BCGs in $\sim 30\%$ of the most overdense regions. We may probe the different populations of different redshifts, and there is no guarantee that our proto-BCGs are on the same evolutionary track towards local BCGs. We do not exclude AGNs from our proto-BCG sample, although the fraction is unlikely to be significant. Also, toy models for major/minor mergers only consider the effect of merger activity, and did not consider its star-formation activity; therefore there is the concern that we cannot adopt these toy models. In any case, our proto-BCG candidates are larger than other field galaxies and other populations, like massive ($M_{*}\sim10^{11}M_{\odot}$) quiescent galaxies at the same redshift.
 \par Again, we argue that this work focus on the UV-brightest galaxies in overdense regions of star-forming galaxies, however, we can not conclude that all BCGs appear from such UV-brightest galaxies at high-redshift. Some sub-millimeter galaxies or massive quiescent galaxies can be the progenitors of BCGs. However, this study shows the different properties of UV-brightest galaxies compared to galaxies in the blank field, and this can be a key to solve the formation and the progenitor of BCGs.
 \begin{figure}[h]
\begin{center}
\includegraphics[width = 90 mm, bb = 0 0 576 432]{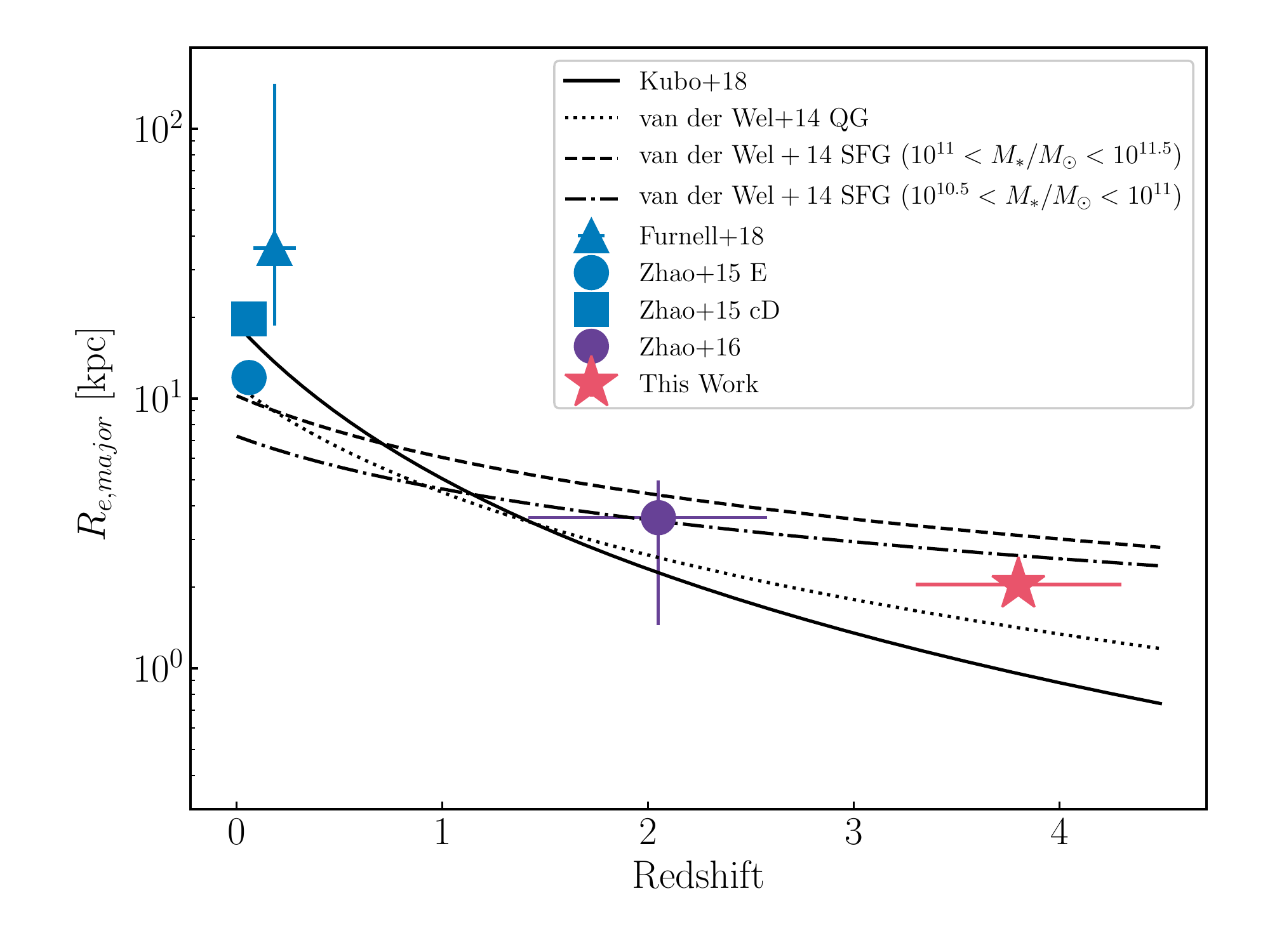}
\caption{The size evolution from proto-BCG candidates to local BCGs at the same stellar mass:$10^{11}<M_{*}/M_{\odot}<10^{11.5}$. The red star represents the size of the median stacked image of proto-BCG candidates. The purple circle is median radii of progenitors of BCGs in \citet{Zhao16}. Their error bar is the 84 and 16 percentiles of the size distribution. Blue circle and square represent BCG radii from \citet{Zhao15}. The Blue triangle represents that from \citet{Furnell18}. The solid black line represents the size evolution track of massive quiescent galaxies from \citet{Kubo18}. The dotted and dashed lines are the size evolution track of quiescent galaxies and star-forming galaxies from \citet{vanderWel14}.}
\end{center}
\end{figure}\label{fig:sizeevo}
\begin{figure}[h]
\begin{center}
\includegraphics[width = 90 mm, bb =  0 0 576 432]{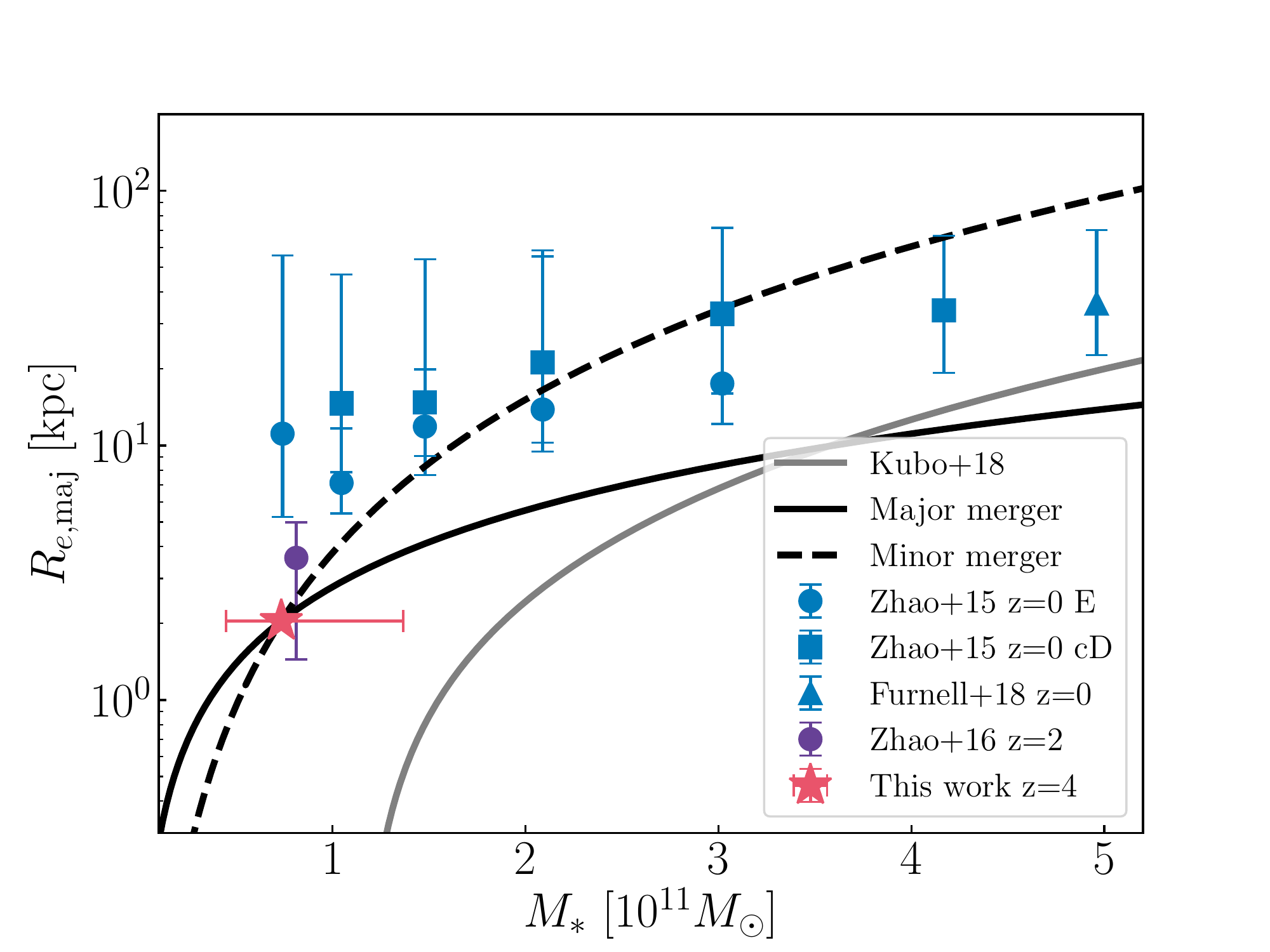}
\caption{The size-mass relation. To compare previous study, we plot BCGs at $z\sim2$ \citep[A purple circle,][]{Zhao16}, $0.05<z<0.3$ \citep[Blue triangles,][]{Furnell18}, $0.05<z<0.1$ \citep[Blue circles and squares,][]{Zhao15}. The solid black line and the dashed line represents the mass growth trend by major merger and minor merger from this work, respectively \citep{Bezanson09, Naab09}. The solid gray line represents the massive quiescent galaxies obtained in \citet{Kubo18}. }
\end{center}
\end{figure}\label{fig:sizemass}

\section{Conclusion}
We carried out a statistical analysis of the UV-brightest galaxies in protoclusters, which is likely to be progenitors of Brightest Cluster Galaxies (proto-BCGs) at $z\sim4$ based on the 179 protocluster candidates identified from the HSC-SSP survey \citep{Toshikawa18}. \par
\begin{enumerate}
\item We constructed a clean sample of $g$-dropout galaxies and identified 63 proto-BCG candidates, which we defined as the brightest objects whose $i$-band magnitudes are $> 1$ mag brighter than the fifth brightest galaxy in each protocluster.
\item We compared the rest-UV color $(i-z)$ of our proto-BCG candidates and field galaxies. We found the proto-BCGs to be redder than field galaxies in the rest-UV. According to the Anderson-Darling test, the difference in color distribution between these two samples was significant.
\item We compared the rest-UV color of protocluster members and field galaxies. Members of protoclusters hosting proto-BCGs are redder than field galaxies. On the other hand, the color distributions of protoclusters without proto-BCGs is the same as that of field galaxies. This indicates that galaxies in the overdense regions around proto-BCGs contain more dust than other star-forming galaxies. We interpret this as meaning that they have experienced early star formation or starbursts. Furthermore, the observed enhancement of bright galaxies indicates that further active star formation is likely.
\item We derived the average radial profiles of proto-BCG candidates and field galaxies by applying the stacking method. We evaluated the effective radii of the stacked images by using {\tt GALFIT} and found the candidate proto-BCG to have larger effective radius than field galaxies. We also estimated the effective radii while taking the effect of dust into account. The difference in the amount of the dust could not explain all of the differences in size between the two samples. We compared the sizes of the BCGs and massive quiescent galaxies at different redshift. Based on the size-stellar mass growth and the toy model, we suggest that our proto-BCGs mainly evolve into local BCGs via minor mergers.
\end{enumerate}
\par Our protocluster candidates have not yet been spectroscopically confirmed. Although these protocluster candidates are likely to evolve into massive clusters at the present day, follow-up spectroscopic observations are required to remove fore/background galaxies from the candidate protocluster members. We are currently surveying protoclusters at $z\sim 2-6$ using the HSC-SSP data and the method of \citet{Toshikawa18}. This ongoing survey will enable us to select proto-BCGs at different redshifts and to track the evolution of the proto-BCGs over $z\sim 2-6$. As the HSC-SSP survey is still ongoing, we will obtain more candidates at  $z\sim4$. We will carry out improved statistical analysis of the properties of protoclusters and their BCGs in the near future.
\section*{Acknowledgements}
\par We thank Tadafumi Takata, Daisuke Iono, Kouichiro Nakanishi for the helpful comments of this research during the thesis defense of KI. Also, we appreciate with the anonymous referee for helpful comments and suggestions that improved the manuscript.
\par This work is based on data collected at the Subaru Telescope and retrieved from the HSC data archive system, which is operated by the Subaru Telescope and Astronomy Data Center at the National Astronomical Observatory of Japan.
\par This work was partially supported by Overseas Travel Fund for Students (2018) of the Department of Astronomical Science, SOKENDAI (the Graduate University for Advanced Studies). NK acknowledges support from the JSPS grant 15H03645. RAO is grateful for financial support from FAPERJ, CNPq and FAPESP. \par  The Hyper Suprime-Cam (HSC) collaboration includes the astronomical communities of Japan and Taiwan and Princeton University.  The HSC instrumentation and software were developed by the National Astronomical Observatory of Japan (NAOJ), the Kavli Institute for the Physics and Mathematics of the Universe (Kavli IPMU), the University of Tokyo, the High Energy Accelerator Research Organization (KEK), the Academia Sinica Institute for Astronomy and Astrophysics in Taiwan (ASIAA), and Princeton University.  Funding was contributed by the FIRST program from the Japanese Cabinet Office, the Ministry of Education, Culture, Sports, Science and Technology (MEXT), the Japan Society for the Promotion of Science (JSPS),  Japan Science and Technology Agency  (JST),  the Toray Science  Foundation, NAOJ, Kavli IPMU, KEK, ASIAA,  and Princeton University.
\par  The Pan-STARRS1 Surveys (PS1) have been made possible through contributions of the Institute for Astronomy, the University of Hawaii, the Pan-STARRS Project Office, the Max-Planck Society, and its participating institutes, the Max Planck Institute for Astronomy, Heidelberg and the Max Planck Institute for Extraterrestrial Physics, Garching, The Johns Hopkins University, Durham University, the University of Edinburgh, Queen's University Belfast, the Harvard-Smithsonian Center for Astrophysics, the Las Cumbres Observatory Global Telescope Network Incorporated, the National Central University of Taiwan, the Space Telescope Science Institute, the National Aeronautics and Space Administration under Grant No. NNX08AR22G issued through the Planetary Science Division of the NASA Science Mission Directorate, the National Science Foundation under Grant No. AST-1238877, the University of Maryland, and Eotvos Lorand University (ELTE).
\par This paper makes use of software developed for the Large Synoptic Survey Telescope. We thank the LSST Project for making their code available as free software at \url{http://dm.lsst.org}.
\bibliographystyle{aasjournal}

\appendix
{
Here, we summarize the method to derive the error of the radial profile of a stacked image obtained at Section \ref{sec:mor} and Section \ref{sec:mordust}. The basis is the same as \citet{Momose14}.}\\
\begin{enumerate}
\item {We obtain about 10000 sky images of HSC $i$-band data. Here, we define the sky image as the image that has no object in $8\ {\rm arcsec}$ from the image center. }
\item {We stack random selected sky images in the same way as we conduct to object images in Section \ref{sec:mor}. The number of sky images is the same as the number of images that we stacked for the radial profile (i.e., 63 images for the radial profile for proto-BCGs and 628 images for that of field galaxies). We make stacked images 1000 times for each radial profile.}
\item {We make the radial profile for each stacked image. From the distribution of each bin's value, we estimate the $1\sigma$ value of the distribution.}
\item {We assume the $1\sigma$ value as the error of the bin of the radial profile.}
\end{enumerate}

\end{document}